Working Paper

# Margin trading, short selling and corporate green innovation

*Ge-zhi Wu[1,2] | Da-ming You[1,2]*

**KEYWORDS:** Margin trading, The quantity of green innovation, The quality of green innovation, Short selling, Corporate green innovation.

[1] Business School, Central South University, Changsha, China

[2] Collaborative Innovation Center of Resource-conserving and Environmentally friendly Society and Ecological Civilization, Central South University, Changsha, China

**Correspondence**

Ge-zhi Wu, Business School, Central South University; Collaborative Innovation Center of Resource-conserving and Environmentally friendly Society and Ecological Civilization, Central South University, No. 932 Lushan South Road, Yuelu District, Changsha, Hunan 410083, China
Personal Email: 649792656@qq.com
Institutional Email：wgzz2218@csu.edu.cn
Telephone number: 86-18975833865

**Funding information**
National Natural Science Foundation of China, Grant/Award Numbers: 71974209;

**ACKNOWLEDGMENTS**
The authors acknowledge the anonymous reviewers and editors for helpful guidance on prior versions of the article.

**ABSTRACT**

This paper uses the panel data of Chinese listed companies from 2007 to 2019, uses the relaxation of China's margin trading and short selling restrictions as the basis of quasi experimental research, and then constructs a double difference model to analyze whether the margin trading and short selling will encourage enterprises to engage in green technology innovation activities. Firstly, our research results show that after the implementation of the margin trading and short selling, the green technology innovation behavior of pilot companies will increase significantly. We believe that the short selling threat and pressure brought by short selling to enterprises are the main reasons for pilot enterprises to engage in green technology innovation. Secondly, the empirical results show that the implementation of margin trading and short selling will significantly promote the quantity of green technology innovation of pilot enterprises, but will not significantly promote the quality of green technology innovation of pilot enterprises. Furthermore, we analyze the difference of the impact of margin trading and short selling on the quantity of green technology innovation of pilot enterprises in different periods. Finally, we find that the performance decline, yield gap between financial assets and operating assets, the risk of stock price crash, management shareholding, institutional shareholding ratio, product market competition, short selling intensity, margin trading intensity and formal environmental regulation intensity will affect the role of policy in promoting green technology innovation of pilot enterprises.

# 1| INTRODUCTION

Sustainable development is an important issue of global concern. Countries all over the world are making efforts to promote green development. When facing the problem of pollution prevention and control, enterprises, as a micro subject, are an important participant. Green technology innovation is an important factor to promote the win-win development of environment and economy. However, because green technology innovation inherently has the dual external attributes of environment and technology, how to encourage enterprises to carry out green technology innovation is a crucial problem.

COVID-19, which broke out in 2020, caused the collapse of the global stock market. Many governments adopted restrictions on margin trading and short selling to try to stabilize and prevent stock prices from falling too much due to panic, the role and influence of margin and short selling mechanism have been paid more and more attention and discussed by stakeholders. Capital market is an important factor affecting micro enterprise behavior, especially enterprise investment behavior (Barro, 1990; Morck, 1990), in which understanding the economic consequences of short selling policy is very important in financial research. The research on this topic is from the impact on short selling nature(Irvine et al,2007; Blau and Wade,2012), Stock market efficiency and pricing (Danielsen and sorescu, 2001; Hirshleifer etal., 2011；Saffi and Sigurdsson, 2011;Ge, 2016)to discussed the impact of short selling on corporate behavior and choice from the perspective of Corporate external governance (Massa et al., 2015;Fang et al., 2016; Chang and Lin et al.,2014, 2015 and 2019；Hope et al, 2017), however, there is still no conclusion on whether and how the stock market margin trading and short selling policy affects enterprise green technology innovation. Therefore, by analyzing how the margin trading and short selling policy affect the company's green technology innovation behavior, we provide clues for the realization of green development.

Unlike most developed countries, margin trading and short selling in China's securities market are gradually implemented on a pilot basis in batches. This gradual expansion provides quasi natural experimental conditions for empirical research and the cornerstone of our research. This study calculates the green technology innovation level of target enterprises and non-target enterprises in china from 2007 to 2019, as shown in Figure 1.Intuitively, since the implementation of the margin trading and short selling in 2010, especially after the large-scale expansion of the underlying securities in 2013, the upward trend of green technology innovation per year of the target enterprises of margin trading and short selling is significantly stronger than that of non-target enterprises. Will the margin trading and short selling have an impact on the green technology innovation behavior of entity enterprises?

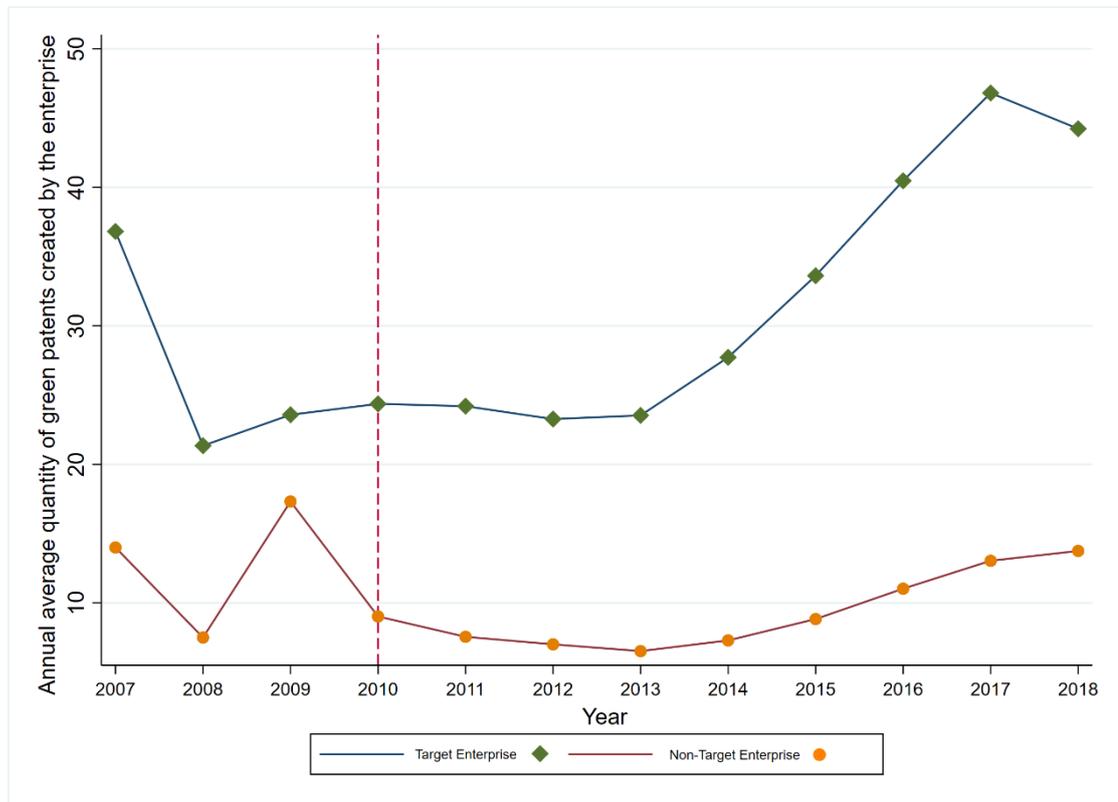

**Figure 1 Change trend of green technology innovation of target enterprises and non-target enterprises**

As Engelberg et al.(2012) points out, short sellers not only have information advantages, but also have rich trading ability and experience. They participate in the formulation and implementation of short selling plan, which poses a great threat to the company. In recent years, as green development has been valued by governments all over the world, in order to deal with the threat of short selling, enterprises have the motivation to transform the environmental pressure they face into green technology innovation. Therefore, we believe that the relax of short selling restrictions in china can promote enterprise green technology innovation.

Firstly, using panel data from Chinese listed companies, we construct a double difference model to analyze whether margin trading and short selling will encourage enterprises to engage in green technology innovation activities. results show that the implementation of margin trading and short selling will significantly promote the quantity of green technology innovation, but will not significantly promote the quality of green technology innovation.

Secondly, this paper also contributes to the literature on the driving factors of green innovation by studying the relationship between margin trading, short selling and enterprise green technology innovation.

Third, we found that the performance decline, yield gap between financial assets and operating assets, the risk of stock price crash, management shareholding, institutional shareholding ratio, product market competition, short selling intensity, margin trading intensity and formal environmental regulation intensity will affect the role of short selling in promoting green technology innovation.

Fourth, the research results of this paper also have important policy significance. These are issues of great concern to decision makers, because how to make enterprises bear the green technology innovation behavior with certain risks and dual externalities is an important issue. We find that the margin trading and short selling mechanism is more similar to a supplementary mechanism spontaneously generated by the market. This paper reveals that the short selling mechanism has an important incentive effect on green technology innovation, and some policy suggestions are provided.

## 2 | RELATED LITERATURE AND THE HYPOTHESIS DEVELOPMENT

### 2.1 RELATED LITERATURE

#### 2.1.1 Institutional background

In October 2005, the CSRC(China Securities Regulatory Commission) issued the regulations on administrative supervision of securities companies, allowing securities companies to engage in margin trading and short selling, laying a foundation for the implementation of follow-up policies. China's margin trading and short selling officially came into effect on March 31 2010. At that time, the largest 90 A-share stocks are allowed by the CSRC to conduct margin trading and short selling. These 90 stocks are constituent stocks of the Shanghai and Shenzhen stock indexes. From 2011 to 2019, the number of target enterprises gradually increased to 278, 500, 700, 900, 950 and 1600.

#### 2.1.2 Short selling

##### 2.1.2.1 The nature of short selling

At present, there are still disputes about the nature of short selling. Some scholars believe that short selling is an informed transaction, while others believe that short selling is a speculative transaction. On the one hand, relevant scholars have provided evidence that short selling is an informed transaction (Daske et al.,2005; Irvine et al., 2007; Christophe et al., 2010 and Nefedova, 2012). Christophe et al. (2010) research shows that analysts bought a large number of short positions before downgrade. Meng et al. (2017) reported that Chinese analysts also had short behavior before the rating downgrade; On the other hand, relevant studies also provide evidence that short selling is speculative trading(Blau & Wade, 2012;Blau & Pinegar,2013). Blau & Wade (2012) determined that short sellers will increase short selling before analysts' ratings are downgraded or upgraded. The evidence of inconsistent rating behavior between short sellers and analysts reflects the existence of speculative trading.

**2.1.2.2 Impact of short selling**

The first part of the literature studies whether short selling causes excessive decline of stock price in adverse circumstances, or whether short selling improves information and pricing efficiency. On the one hand, some regulators and enterprise management believe that short selling puts pressure and turblence on the downward trend of stock price, because there is a huge amount of short selling before the huge negative rate of return of the enterprise(Fox et al. 2010). To et al. (2018) also proved that insider traders use short selling mechanism to obtain rent based on information advantage.

On the other hand, Miller (1977), Bris et al. (2007) and Saffi and sigurdsson (2011) believe that short selling reflects and diffuses sufficient negative information, which is beneficial to stock pricing, overall improved pricing efficiency and quality, effectively avoids the huge fluctuation of the stock price in the face of bad news, and promotes the stable adjustment of the stock price. Alves et al. (2016) used the short selling ban in Europe in 2011 to prove that short selling restrictions will improve stock price volatility. Research by Chmura et al. (2019) shows that short selling reduces the volatility of stock returns.

In addition, other scholars have also studied the positive and negative effects of short selling plan on stock market efficiency and pricing (Diamond and Verrecchia, 1987; Allen and Gale，1991). Mainly around stock pricing (BOEHMER and Wu, 2013; Chang and Luo et al., 2014), stock return forecast, market crash and abnormal accrued profit (Hong and Stein, 2003; Hirshleifer etal.，2011；Ge etal.，2016).

The second part studies the relationship between short selling policy and micro corporate behavior. Massa et al. (2015) considered that short selling restricts managers' immoral behavior and their earnings management. Similarly, Chang and Lin et al. (2014, 2015 and 2019) proved that short selling inhibits the establishment of internal alliances and ineffective investment behavior, and confirmed the governance role of short sellers. He and Tian (2016) pointed out that fewer restrictions on short selling promoted the technological innovation of enterprises. Deng and Gao (2018) proved that short selling is positively related to the government punishment faced by enterprises. Karpoff and Lou (2010), short selling plan helps to detect corporate misconduct. Fang et al. (2016) found short selling helps to detect corporate fraud. Meanwhile, De Angelis et al. (2017) discussed the effect of short selling on executive incentive contract and found that it has a positive impact on the granting of executive stock options by enterprises. Hope et al. (2017) found that companies with high short selling threat face higher auditor fees. In addition, Chang et al. (2019) discussed that acquirers facing higher short selling threat have relatively higher performance in M & A. In conclusion, these studies show that short selling has an impact on corporate behavior, but is unclear whether short selling will affect green technology innovation.

**2.1.3 Margin trading**

First, financing transactions strengthen the volatility of stock prices and enhance the short-sighted behavior of managers. Previous studies have found that the leveraged trading nature of financing encourages speculative trading, which is easy to cause the company's stock price to be high, and the managers also tends to maintain the stock price, mainly due to avoid the shrinkage of shareholder wealth, M &A risk (Stein, 1988), the attention of stakeholders such as stock analysts and the media (Hong et al., 2000), which will have an adverse impact on the reputation and career of the managers. Moreover, managers may also maintain the return of its own stock options (Chen et al., 2015a; Edmans et al., 2017) therefore, the company's managers may sacrifice the company's long-term interests to maintain the stock price, or even participate in speculation to meet the needs of stakeholders.

In addition, financing transactions provide conditions for manager's short-term behavior(Pownall and simko, 2005) financing transactions are often speculative (Hardouvelis and Peristiani, 1992). Financing transactions will not improve stock pricing efficiency and increase the content of stock price information, but will reduce the content of individual stock characteristic information(Chang et al., 2014). if the company characteristic information cannot be integrated into the stock price, the ability of stock price to reflect the company's fundamentals will be reduced, at this time, the company's stakeholders will not be able to motivate the managers through mechanisms such as stock options, and the improvement of the company's information environment and asymmetry is not conducive to the stakeholders' supervision of corporate behavior (Benmelech et al., 2010). At this time, the behavior of the managers may also tend to be short-term.

**2.1.4 Green technology innovation**

Many scholars have studied the driving mechanism of green technology innovation from many aspects. The first kind of literature studies from within the enterprise, and the research topics include: enterprise strategy and sustainable development orientation (Claudiy et al., 2016), environmental best management practices (Christmann et al., 2000), CEO and executive expectations, cognition and motivation (Bendell, 2017) and green intellectual capital (Chen, 2008). The second kind of literature starts from the relationship between enterprises, and the research topics include: the impact of green supply chain on green technology innovation (Leoncini et al., 2016), corporate governance and stakeholder impact (Juntunen et al., 2019; Scandelius et al., 2018), external integration capability and knowledge chain of the company (Dangelico et al., 2013). The third kind of literature starts from institutional factors, and its topics include: resources, institutional factors and media pressure (Bansal, 2005), environmental policy (Cainelli et al., 2020; Berrone et al., 201；Holweg, 2014), informal system and policy (Lim et al., 2014). In addition to the above three categories, there are also studies from a transnational perspective, such as the impact of pollution transnational spillover and technology transnational spillover on green technology innovation (BU et al., 2016; Li et al., 2017). At present, there is no research focusing on short selling mechanism and enterprise green technology innovation.

## 2.2 HYPOTHESIS DEVELOPMENT

### 2.2.1. Short selling and green technology innovation

The relax of short selling restrictions can curb the short-sighted behavior of management through external governance and strengthening the effectiveness of Internal Governance: on the one hand, short sellers are usually regarded as informed traders (BOEHMER et al, 2008;

Desai et al., 2002), the existence of short sellers accelerates the spread of negative news (karpoff and Lou, 2010; Young, 2016).Therefore, short selling provides an effective external governance mechanism to supervise the management. If short sellers find the short-term opportunistic tendency of management, it will cause a series of negative chain reactions. Therefore, according to the direct governance effect of short selling mechanism, it helps to curb the short-sighted behavior of management.

On the other hand, short selling mechanism also provides incentives and conditions for the company to strengthen internal governance. The stock price decline caused by short selling will affect the wealth placement of shareholders, and shareholders have a stronger motivation to supervise the company's management, so as to alleviate the agency problems of shareholders and management, reduce agency costs and improve business performance, which is helpful for stakeholders to supervise the management and implement the equity incentive salary design mechanism.

When considering environmental issues and responsibilities, relevant managers are faced with the system and trade-offs of potential strategies, and short sellers will affect the parameters of strategy formulation. Specifically, short selling helps to find corporate environmental default. The emergence and implementation of short selling plan will make managers bear the corresponding economic consequences. Once a company's environmental default is found, it will face greater pressure of short selling when more shares can be lent. Therefore, the company should not only bear the cost of default, but also bear the economic losses caused by the decline of share price. The stock price decline caused by short selling will have many adverse effects on the manager's career, reputation and the company's reputation. Therefore, short selling plays an external governance role for the management. Therefore, the relax of short selling restrictions may have a positive external incentive effect on management's environmental governance and encourage enterprises to engage in green technology innovation.

We expect that when facing the potential short selling threat, the enterprise management will pay more attention to the investment and development of environmental undertakings, and finally turn the environmental pressure caused by short selling threat into green technology innovation. Short selling threat will strengthen the environmental responsibility of enterprises and urge enterprises to increase investment in green technology innovation.

**2.2.2. Margin trading and green innoavtion**

From the perspective of margin trading, margin trading may change management's investment preference and form short-term behavior. Specifically, the proportion of institutional investors in China's stock market is relatively low, and there are a large number of retail investors with short-term transactions in the market. The leverage of financing transactions enlarges the high volatility of the stock market and improves investors' speculation. The stock price of the target enterprise may rise rapidly. In order to maintain the company's stock price, the investment behavior of the management will also change. In addition, due to the impact of performance appraisal and reputation, in order to prevent the sharp fluctuation of stock price, when the management is under the pressure of maintaining falsely high stock price, the management may focus on the improvement of short-term performance, that is, the margin trading may inhibit the enterprise's green technology innovation behavior.

Because margin trading and short selling have different effects on green technology innovation, the effect of policy implementation of relaxing the restrictions on margin trading and short selling depends on the size of two forces, and put forward the hypothesis H1 and H2:

H1: When the effects of short selling is greater than margin trading, the implementation of relaxing the restrictions on margin trading and short selling can promote green technology innovation activities.

H2: When the effects of short selling is less than the margin trading, the implementation of relaxing the restrictions on margin trading and short selling can not promote green technology innovation activities.

## 3| SAMPLE DESCRIPTION AND VARIABLE MEASUREMENT

### 3.1 Data source and sample selection

This study collects relevant data from the databases of the China Stock Market & Accounting Research Database (CSMAR) and Chinese Research Data Services (CNRDS), and uses the main financial indicators and data listed on the Chinese A-share market from 2007 to 2019 as the initial sample. In the data collection process, we try our best to ensure that the sample size is maximized. We started with the implementation of the new accounting standards in 2007. In order to ensure the validity of the data, this paper carries out the following processing: (1) Eliminate the enterprises belonging to the financial industry and insurance industry. (2) Excluding the sample of companies with performance losses which net profit less than 0. (3) Eliminate the enterprises that are transferred out after being included in the list of the subject matter of margin trading and short selling. (4) Eliminate samples with missing and abnormal data. After the above processing, this paper finally obtains 6080 effective observations from 1473 companies.

In Table 1, we have listed the sample distribution by year.

**Table1 Sample screening and distribution by year.**

**Panel A：Sample screening**

|  | observations |
|---|---|
| Initial sample | 8182 |
| Excluding: |  |
|     Eliminate the enterprises belonging to the financial industry and insurance industry | 168 |
|     Excluding the sample of companies with performance losses which net profit less than 0 | 748 |
|     Eliminate the enterprises that are transferred out after being included in the list of the subject matter of margin trading and short selling | 176 |
|     Eliminate samples with missing and abnormal data | 1010 |
| Final sample | 6080 |

**Panel B：Sample distribution by year**

| year | Freq. | Percent | Cum. |
|---|---|---|---|
| 2007 | 17 | 0.28 | 0.28 |
| 2008 | 34 | 0.56 | 0.84 |
| 2009 | 67 | 1.10 | 1.94 |
| 2010 | 145 | 2.38 | 4.33 |
| 2011 | 240 | 3.95 | 8.27 |
| 2012 | 377 | 6.20 | 14.47 |
| 2013 | 456 | 7.50 | 21.97 |
| 2014 | 533 | 8.77 | 30.74 |
| 2015 | 597 | 9.82 | 40.56 |
| 2016 | 799 | 13.14 | 53.70 |
| 2017 | 942 | 15.49 | 69.19 |
| 2018 | 1021 | 16.79 | 85.99 |
| 2019 | 852 | 14.01 | 100 |
| Total | 6080 | 100 |  |

### 3.2| Variable description

### 3.2.1 Green Innovation

1. Quantity index *NGInnovation*

Taking into account that the company's green patent information disclosure may have been disclosed in advance during the research and development process, we use the total number of green invention patents and utility model patent applications of listed companies in the year as a proxy variable for the number of green technology innovation indicators, and we take the logarithm of this variable to solve the problem of skew distribution.

2. Quality Index *QGInnovation*

In order to measure the impact of green patents, we pay attention to the index of the number of citations of green patents as a key indicator for measuring influence. Because the impact of green patents may take several years to officially decline, considering the role of green patents, the total number of green patent applications and the number of cumulative green patent citations, the establishment of indicators to measure the quality of the company's green patents is as follows[1]:

$$\text{Green patent quality}_{i,t} = \frac{\sum_{1996}^{t}\text{Total number of green patents cited}_{i,t}}{\sum_{1996}^{t}\text{Number of green patent applications}_{i,t}}$$

---

[1] Green patent data began to be recorded in 1996.

### 3.3| The benchmark model

Since the batch expansion of the underlying securities of margin trading and short selling has formed a quasi natural experiment, we construct the following multi-stage double difference model to test the impact of margin trading and short selling on enterprise green innovation. Model（2）-（3）as follow,

（2）
$$LN(NGInnovation_{i,t}) = \beta_0 + \beta_1 PostList_{i,t} + \beta_2 List_{i,t} + \sum \beta_j \left(Control\ variables_{i,t}\right) + \sum INDUSTRYID + \sum Year + \varepsilon_{i,t}$$

（3）
$$QGInnovation_{i,t} = \beta_0 + \beta_1 PostList_{i,t} + \beta_2 List_{i,t} + \sum \beta_j \left(Control\ variables_{i,t}\right) + \sum INDUSTRYID + \sum Year + \varepsilon_{i,t}$$

The variable of interest, $LN(NGInnovation_{i,t})$ and $QGInnovation_{i,t}$ measures the quantity and quality of the green innovation of company i in year t; A larger $\beta_1$ represents a greater impact of the margin trading and short selling; $PostList_{i,t}$ refers to the enterprise after being officially included in margin trading and short selling; $List_{i,t}$ refers to the enterprise included in margin trading and short selling.

The control variables mainly include the micro characteristic variables of the enterprise. In addition, we use the ratio of provincial government environmental protection financial expenditure to GDP to control the impact of environmental regulation, and we use the GDP of the province where the listed company is located to control regional factors. See Appendix A for the specific definition of the variables.

## 4| EMPIRICAL RESULTS

### 4.1| Descriptive statistics

This table 2 reports descriptive statistics of green innovation and control variables for the sample 2007–2019.

**Table 2　Decriptive statistics results**

|  | count | mean | sd | min | p50 | max |
|---|---|---|---|---|---|---|
| *NGInnovation* | 6080 | 21.614 | 81.666 | 1.000 | 6.000 | 1869.000 |
| QGInnovation | 6080 | 0.841 | 1.315 | 0.000 | 0.515 | 44.600 |
| List | 6080 | 0.500 | 0.500 | 0.000 | 1.000 | 1.000 |
| RD | 6080 | 0.026 | 0.024 | 0.000 | 0.021 | 0.272 |
| GDP | 6080 | 10.528 | 0.661 | 6.934 | 10.506 | 11.587 |
| Lev | 6080 | 0.456 | 0.185 | 0.048 | 0.460 | 0.885 |
| Growth | 6080 | 0.207 | 0.353 | -0.467 | 0.138 | 2.694 |
| Tan $qi$ | 6080 | 0.920 | 0.087 | 0.520 | 0.950 | 1.000 |
| Cfo | 6080 | 0.048 | 0.060 | -0.156 | 0.044 | 0.254 |
| Board | 6080 | 2.162 | 0.199 | 1.609 | 2.197 | 2.708 |
| Indep | 6080 | 0.374 | 0.055 | 0.273 | 0.333 | 0.625 |
| Top1 | 6080 | 0.343 | 0.153 | 0.072 | 0.321 | 0.797 |
| Soe | 6080 | 0.399 | 0.490 | 0.000 | 0.000 | 1.000 |
| ER | 6080 | 0.190 | 0.070 | 0.087 | 0.176 | 1.379 |
| Roa | 6080 | 0.051 | 0.041 | 0.000 | 0.042 | 0.238 |
| Roe | 6080 | 0.094 | 0.070 | 0.000 | 0.081 | 0.421 |
| ATO | 6080 | 0.679 | 0.401 | 0.056 | 0.590 | 2.907 |
| Dual | 6080 | 0.243 | 0.429 | 0.000 | 0.000 | 1.000 |
| BM | 6080 | 1.168 | 1.210 | 0.051 | 0.748 | 8.232 |
| TobinQ | 6080 | 1.910 | 1.113 | 0.815 | 1.565 | 17.676 |
| Size | 6080 | 299.238 | 1172.731 | 4.152 | 61.565 | 24049.100 |
| ListAge | 6080 | 2.252 | 0.669 | 0.693 | 2.303 | 3.332 |
| FirmAge | 6080 | 2.834 | 0.347 | 0.693 | 2.890 | 3.555 |
| Balance | 6080 | 0.702 | 0.586 | 0.018 | 0.537 | 2.961 |
| Dturn | 6080 | -0.094 | 0.417 | -2.494 | -0.039 | 1.585 |
| INST | 6080 | 0.429 | 0.232 | 0.000 | 0.443 | 0.889 |
| Mshare | 6080 | 0.116 | 0.176 | 0.000 | 0.006 | 0.705 |
| Mfee | 6080 | 0.088 | 0.064 | 0.007 | 0.074 | 0.678 |
| Occupy | 6080 | 0.015 | 0.019 | 0.000 | 0.009 | 0.202 |
| Big 4 | 6080 | 0.086 | 0.280 | 0.000 | 0.000 | 1.000 |

**4.2| Multivariate analysis**

**4.2.1 Margin trading, short selling and the quantity of green technology innovation**

In order to test the influence of margin trading and short selling on the quantity of green technology innovation more precisely, this paper makes a multivariate regression analysis. The regression results are shown in columns (1) of table 3 The results show that the coefficient of *PostList* is positive significance at 1% confidence level. The implementation of margin trading and short selling will promote the green technology innovation in general. From the perspective of each control variable, the regression results are basically in line with the expectation. Further, in columns (2) to (6) of table 3, the policy has a significant role in promoting quantity of green technology innovation in different period. (In the time phased study, the sample data of 2010, 2012, 2013, 2014 and 2017 are excluded respectively to exclude the interference of the impact of the expansion year)

**Table 3-1 Margin trading, short selling and the quantity of green technology innovation**

| Sample time interval | Full sample | 2007-2013 excluding 2010 | 2008-2016 excluding 2012 |
|---|---|---|---|
| | （1） | （2） | （3） |
| Dependent variable | $LN(NGInnovation)$ | $LN(NGInnovation)$ | $LN(NGInnovation)$ |
| *PostList* | 0.3449*** | 0.4981*** | 0.3487*** |
| | (4.6394) | (3.5499) | (4.2669) |
| *List* | 0.0861 | 0.1894* | 0.1361 |
| | (1.0233) | (1.7489) | (1.4992) |
| *RD* | 4.0429*** | 0.9839 | 2.8806 |
| | (3.3060) | (0.3908) | (1.4694) |
| *GDP* | 0.0802 | 0.0538 | 0.1270 |
| | (1.1101) | (0.3626) | (1.3830) |
| *Lev* | 0.6663*** | 0.9597** | 0.2128 |
| | (3.2375) | (2.4172) | (0.7901) |
| *Growth* | 0.0695 | 0.0435 | 0.0779 |
| | (1.4698) | (0.3900) | (1.1506) |
| $\tan qi$ | 0.3691 | 1.4985* | 0.6007* |
| | (1.4578) | (1.7355) | (1.7876) |
| *Cfo* | -0.3706 | -0.5090 | -0.4915 |
| | (-1.0252) | (-0.7207) | (-0.9807) |
| *Board* | 0.0239 | -0.0332 | 0.0067 |
| | (0.1414) | (-0.1176) | (0.0311) |
| *Indep* | 0.3741 | 0.2111 | 0.6039 |
| | (0.7223) | (0.2116) | (0.8890) |
| $Top1$ | -0.3089 | -0.3183 | -0.5086 |
|       | (-1.0911)  | (-0.6786)  | (-1.4855)  |
| Soe   | 0.0791     | -0.0142    | 0.0240     |
|       | (0.9764)   | (-0.1148)  | (0.2425)   |
| ER    | 0.1508     | 0.3693     | 0.0631     |
|       | (0.2831)   | (0.2653)   | (0.0937)   |
| Roa   | -0.8775    | -0.0405    | -4.3409*   |
|       | (-0.4650)  | (-0.0125)  | (-1.7089)  |
| Roe   | 2.2386**   | 1.4153     | 4.1734***  |
|       | (2.0224)   | (0.8090)   | (2.7354)   |
| ATO   | -0.0405    | 0.0847     | -0.0278    |
|       | (-0.4271)  | (0.6665)   | (-0.2378)  |
| Dual  | 0.0034     | 0.0388     | -0.0667    |
|       | (0.0608)   | (0.3529)   | (-0.9453)  |
| BM    | 0.0749**   | 0.0135     | 0.0703     |
|       | (2.1012)   | (0.2136)   | (1.2421)   |
| TobinQ| -0.1429*** | -0.0618    | -0.1201*** |
|       | (-6.2233)  | (-0.8194)  | (-4.7249)  |
| Size  | 0.0002***  | 0.0003***  | 0.0002***  |
|       | (4.7837)   | (4.7484)   | (3.7688)   |
| ListAge | -0.0335  | -0.0683    | -0.1330*   |
|       | (-0.5515)  | (-0.6752)  | (-1.6824)  |
| FirmAge | -0.2599*** | -0.2137  | -0.2101*   |
|       | (-2.5838)  | (-1.4825)  | (-1.7740)  |
| Balance | 0.0703   | 0.0870     | 0.0418     |
|       | (1.1358)   | (0.7713)   | (0.5004)   |
| Dturn | 0.0809**   | -0.0509    | 0.0870     |
|       | (2.0499)   | (-0.5040)  | (1.4069)   |
| INST  | 0.5111***  | 0.1703     | 0.4098**   |
|       | (3.9124)   | (0.7328)   | (2.4846)   |
| Mshare| 0.0211     | -0.2965    | -0.2950    |
|       | (0.1240)   | (-0.9611)  | (-1.3645)  |
| Mfee  | -0.5573    | 0.7450     | -0.2967    |
|       | (-1.0897)  | (0.6663)   | (-0.4375)  |
| Occupy| 0.8744     | -0.0300    | 2.7095*    |
|       | (0.7523)   | (-0.0110)  | (1.6951)   |
| Big4  | 0.4564***  | 0.5654***  | 0.4947***  |
|       | (3.2122)   | (3.2181)   | (2.9728)   |
| _cons | 0.8027     | -0.0972    | 0.1252     |
|       | (0.7586)   | (-0.0422)  | (0.0965)   |
| Industry | Yes     | Yes        | Yes        |
| Year  | Yes        | Yes        | Yes        |
| N     | 6080       | 1191       | 2871       |
| adj. $R^2$ | 0.318 | 0.312      | 0.317      |

$t$ statistics based on robust standard error in parentheses $^{*} p < 0.1$, $^{**} p < 0.05$, $^{***} p < 0.01$

**Table 3-2 Margin trading, short selling and the quantity of green technology innovation**

| Sample time interval | 2010-2017 excluding 2013 | 2012-2017 excluding 2014 | 2015-2019 excluding 2017 |
|---|---|---|---|
| | （4） | （5） | （6） |
| Dependent variable | $LN(NGInnovation)$ | $LN(NGInnovation)$ | $LN(NGInnovation)$ |
| PostList | 0.3599*** | 0.3142*** | 0.5205*** |
| | (4.3416) | (3.3901) | (2.9298) |
| List | 0.1158 | 0.1596 | -0.1455 |
| | (1.2330) | (1.6281) | (-0.8043) |
| RD | 5.2446*** | 4.5837*** | 4.2744*** |
| | (2.9665) | (2.5849) | (3.4996) |
| GDP | 0.0849 | 0.0739 | 0.0659 |
| | (1.0019) | (0.9024) | (0.8772) |
| Lev | 0.4620* | 0.5293** | 0.4761* |
| | (1.9488) | (2.1320) | (1.9271) |
| Growth | 0.0989* | 0.0999 | 0.0787 |
| | (1.7126) | (1.5827) | (1.2394) |
| $Tan_{qi}$ | 0.3704 | 0.4747* | 0.4515 |
| | (1.3099) | (1.6567) | (1.5788) |
| Cfo | -0.5485 | -0.4880 | -0.1624 |
| | (-1.2002) | (-1.0725) | (-0.3687) |
| Board | 0.0995 | -0.0393 | -0.1781 |
| | (0.5057) | (-0.2087) | (-0.9897) |
| Indep | 0.5523 | 0.2418 | 0.2976 |
| | (0.8754) | (0.3685) | (0.5306) |
| $Top_1$ | -0.5050 | -0.3325 | -0.1904 |
| | (-1.5532) | (-0.9778) | (-0.5603) |
| Soe | 0.0579 | 0.0627 | 0.0869 |
| | (0.6200) | (0.6587) | (1.0194) |
| ER | 0.1933 | 0.1577 | 0.5580 |
| | (0.3191) | (0.2696) | (1.0008) |
| Roa | -2.4622 | -2.5416 | -2.5302 |
| | (-1.1558) | (-1.1294) | (-1.0791) |
| Roe | 3.1270** | 3.3429** | 3.5139** |
| | (2.5456) | (2.5308) | (2.4892) |
| ATO | -0.0565 | -0.0556 | -0.1432 |
| | (-0.5018) | (-0.4760) | (-1.3094) |
| Dual | 0.0064 | 0.0008 | -0.0310 |
| | (0.0992) | (0.0123) | (-0.5012) |
| BM | 0.0924 | 0.0599 | 0.1072*** |

|          | (1)        | (2)        | (3)        |
|----------|------------|------------|------------|
|          | (1.5441)   | (1.2111)   | (3.2159)   |
| TobinQ   | -0.1234*** | -0.1343*** | -0.1548*** |
|          | (-5.0453)  | (-5.5621)  | (-6.5632)  |
| Size     | 0.0002***  | 0.0002***  | 0.0002***  |
|          | (4.0716)   | (4.3003)   | (4.8604)   |
| ListAge  | -0.0958    | -0.0649    | -0.0050    |
|          | (-1.3055)  | (-0.9068)  | (-0.0747)  |
| FirmAge  | -0.2225**  | -0.2728**  | -0.3001*** |
|          | (-1.9825)  | (-2.4237)  | (-2.6788)  |
| Balance  | 0.0585     | 0.0879     | 0.0761     |
|          | (0.7859)   | (1.1550)   | (1.0876)   |
| Dturn    | 0.1242**   | 0.1205**   | 0.0396     |
|          | (2.5136)   | (2.3394)   | (0.7053)   |
| INST     | 0.4537***  | 0.4106**   | 0.5705***  |
|          | (2.9043)   | (2.5365)   | (3.4961)   |
| Mshare   | -0.0771    | 0.0314     | 0.1464     |
|          | (-0.3834)  | (0.1510)   | (0.7572)   |
| Mfee     | -0.9976    | -0.6838    | -0.7991    |
|          | (-1.6172)  | (-1.0114)  | (-1.3326)  |
| Occupy   | 3.2162**   | 3.3424**   | 0.6046     |
|          | (2.3000)   | (2.2270)   | (0.4602)   |
| Big4     | 0.4398***  | 0.4579***  | 0.4259***  |
|          | (2.7165)   | (2.7633)   | (2.8807)   |
| _cons    | 0.4250     | 0.9678     | 1.5611     |
|          | (0.3531)   | (0.8088)   | (1.3933)   |
| Industry | Yes        | Yes        | Yes        |
| Year     | Yes        | Yes        | Yes        |
| N        | 3633       | 3171       | 3269       |
| adj. $R^2$ | 0.323    | 0.320      | 0.321      |

$t$ statistics based on robust standard error in parentheses* $p < 0.1$, ** $p < 0.05$, *** $p < 0.01$

### 4.2.2 Margin trading, short selling and the quality of green technology innovation

In order to more accurately test the impact of margin trading and short selling on the quality of green technology innovation, this paper makes a multiple regression analysis. The regression results are presented in Column (1) of table 4 shows the analysis results of the whole sample. The results show that the coefficient of *PostList* under the full sample is positive, but it is only significant at the 10% confidence level, combined with the robustness test results of the fixed effect model later, the coefficient is not significant at all, indicating that the policy does not significantly improve the quality of green technology innovation. Further, in columns (2) to (6) of table 4, we also show the impact of margin trading and short selling on the quality of enterprise green technology innovation in each expansion stage. From the regression results, although in the third and fourth stages, the promotion effect of policies on the quality of green patents is significant at the 10% confidence level, it is very weak (P values are 0.095 and 0.092, respectively), so overall the impact of margin trading and short selling on the quality of green technology innovation is not significant. On the whole, it shows that the policy is difficult to improve the quality of green technology innovation. From each control variable, the regression results basically meet the expectations. (In the time phased study, the sample data of 2010, 2012, 2013, 2014 and 2017 are excluded respectively to exclude the interference of the impact of the expansion year)

**Table 4-1 Margin trading, short selling and the quality of green technology innovation**

| Sample time interval | Full sample | 2007-2013 excluding 2010 | 2008-2016 excluding 2012 |
|---|---|---|---|
| | （1） | （2） | （3） |
| Dependent variable | *QGInnovation* | *QGInnovation* | *QGInnovation* |
| *PostList* | 0.1364* | -0.0869 | 0.0692 |
| | (1.8492) | (-0.7343) | (1.0282) |
| *List* | -0.0539 | -0.1563 | -0.0592 |
| | (-0.5883) | (-1.0169) | (-0.7389) |
| *RD* | 4.9719** | 4.1950 | 4.9030** |
| | (2.2995) | (1.5132) | (2.2639) |
| *GDP* | 0.0862 | 0.2433* | -0.0129 |
| | (1.2920) | (1.7049) | (-0.1369) |
| *Lev* | -0.3050 | -0.5760* | -0.3072 |
| | (-1.0270) | (-1.6596) | (-0.8906) |
| *Growth* | 0.0552 | 0.3643 | -0.0643 |
| | (0.9622) | (1.1996) | (-1.1204) |
| Tan *qi* | 0.4324 | -0.7259 | 0.4977 |
|         | (1.4227)  | (-0.3449)  | (1.2062)  |
| Cfo     | 0.2379    | 2.4600**   | 0.0989    |
|         | (0.5907)  | (2.1517)   | (0.1935)  |
| Board   | -0.0727   | -0.1096    | 0.0620    |
|         | (-0.5421) | (-0.4452)  | (0.3841)  |
| Indep   | -0.3302   | -1.1382    | 0.1520    |
|         | (-0.6986) | (-0.8402)  | (0.2632)  |
| Top1    | 0.3492    | 1.5679     | -0.0097   |
|         | (0.8123)  | (1.2842)   | (-0.0220) |
| Soe     | -0.0627   | -0.1023    | -0.1424   |
|         | (-0.8634) | (-0.8632)  | (-1.4644) |
| ER      | 0.0386    | 0.2977     | 0.0516    |
|         | (0.2632)  | (1.0847)   | (0.2602)  |
| Roa     | 0.0264    | -4.5613    | -0.3643   |
|         | (0.0070)  | (-1.6092)  | (-0.1105) |
| Roe     | -0.1505   | 2.1046     | 0.5059    |
|         | (-0.0823) | (1.4529)   | (0.3219)  |
| ATO     | -0.2408** | -0.4051*   | -0.2461   |
|         | (-2.0213) | (-1.6946)  | (-1.3616) |
| Dual    | -0.0516   | -0.2366*   | -0.0484   |
|         | (-0.7655) | (-1.7212)  | (-0.4862) |
| BM      | -0.0004   | -0.0383    | 0.0210    |
|         | (-0.0157) | (-0.9227)  | (0.5473)  |
| TobinQ  | 0.0772    | 0.0959     | 0.0628    |
|         | (1.4553)  | (1.6112)   | (1.5943)  |
| Size    | 0.0000**  | 0.0001**   | 0.0000    |
|         | (2.0972)  | (1.9911)   | (1.4859)  |
| ListAge | 0.2416*** | 0.2036     | 0.1904**  |
|         | (2.7619)  | (0.8300)   | (2.2395)  |
| FirmAge | 0.0020    | 0.1333     | -0.0121   |
|         | (0.0179)  | (0.7116)   | (-0.1095) |
| Balance | 0.1278    | 0.5999     | 0.0578    |
|         | (1.0069)  | (1.3798)   | (0.4765)  |
| Dturn   | 0.0148    | 0.0386     | 0.0100    |
|         | (0.3357)  | (0.2246)   | (0.1993)  |
| INST    | -0.2998** | -0.6104    | -0.2430*  |
|         | (-2.0415) | (-1.5798)  | (-1.7390) |
| Mshare  | -0.2965   | -0.7057    | -0.3842   |
|         | (-1.2169) | (-1.4174)  | (-1.2995) |
| Mfee    | 0.3866    | -1.4500    | 0.2933    |
|         | (0.5884)  | (-1.0619)  | (0.3267)  |
| Occupy  | 0.3237    | 3.7511     | -0.9675   |
|         | (0.3108)  | (1.2356)   | (-0.6428) |
| Big4    | 0.0672    | -0.0872    | 0.0968    |

|  | (0.8952) | (-0.6012) | (1.0330) |
|---|---|---|---|
| _cons | -0.9882 | -1.3556 | -0.3516 |
|  | (-1.2029) | (-0.6240) | (-0.3758) |
| Industry | Yes | Yes | Yes |
| Year | Yes | Yes | Yes |
| N | 6080 | 1191 | 2871 |
| adj. $R^2$ | 0.092 | 0.111 | 0.084 |

$t$ statistics based on robust standard error in parentheses $^{*}\,p<0.1$, $^{**}\,p<0.05$, $^{***}\,p<0.01$

**Table 4-2 Margin trading, short selling and the quality of green technology innovation**

| Sample time interval | 2010-2017 excluding 2013 | 2012-2017 excluding 2014 | 2015-2019 excluding 2017 |
|---|---|---|---|
|  | （4） | （5） | （6） |
| Dependent variable | QGInnovation | QGInnovation | QGInnovation |
| PostList | 0.1472* | 0.1854* | -0.0320 |
|  | (1.6691) | (1.6855) | (-0.2210) |
| List | -0.1090 | -0.1104 | 0.1555 |
|  | (-0.9743) | (-0.8095) | (1.0283) |
| RD | 4.9713** | 5.2204** | 5.3929** |
|  | (2.2238) | (2.0593) | (2.0535) |
| GDP | 0.0566 | 0.0640 | 0.0512 |
|  | (0.6674) | (0.7308) | (0.6992) |
| Lev | -0.3632 | -0.3786 | -0.0601 |
|  | (-1.0450) | (-1.0712) | (-0.1542) |
| Growth | 0.0621 | 0.0876 | 0.0067 |
|  | (0.7314) | (0.9722) | (0.1545) |
| Tan $qi$ | 0.4701 | 0.3750 | 0.3021 |
|  | (1.2796) | (1.0050) | (0.9852) |
| Cfo | 0.3684 | 0.0508 | -0.4304 |
|  | (0.6431) | (0.0939) | (-1.0969) |
| Board | -0.0638 | -0.2123 | -0.1166 |
|  | (-0.3658) | (-1.1058) | (-0.7388) |
| Indep | -0.3803 | -0.6323 | -0.2221 |
|  | (-0.6014) | (-0.9060) | (-0.4288) |
| Top1 | 0.4826 | 0.9083 | 0.1025 |
|  | (0.8987) | (1.5084) | (0.2966) |
| Soe | -0.0991 | -0.0235 | 0.0035 |
|  | (-1.0968) | (-0.2750) | (0.0456) |
| ER | 0.0243 | 0.0411 | 0.0122 |
|  | (0.1337) | (0.2135) | (0.0779) |
| Roa | 0.0007 | 0.6127 | 2.0734 |
|  | (0.0002) | (0.1334) | (0.3761) |
| Roe | 0.1123 | 0.2523 | -1.0146 |
| | (0.0572) | (0.1134) | (-0.3789) |
| ATO | -0.2918* | -0.4443** | -0.2358* |
| | (-1.6749) | (-2.2941) | (-1.9483) |
| Dual | -0.0722 | -0.1086 | -0.0459 |
| | (-0.8160) | (-1.3462) | (-0.7050) |
| BM | 0.0419 | 0.0421 | -0.0002 |
| | (0.8615) | (0.9803) | (-0.0093) |
| TobinQ | 0.0948* | 0.0943 | 0.0513 |
| | (1.6949) | (1.5644) | (1.1320) |
| Size | 0.0000 | 0.0000 | 0.0000** |
| | (1.4701) | (1.4900) | (2.0175) |
| ListAge | 0.2399** | 0.2547** | 0.2480*** |
| | (2.1996) | (2.1675) | (3.2182) |
| FirmAge | 0.0514 | 0.0922 | -0.0308 |
| | (0.4122) | (0.6260) | (-0.2618) |
| Balance | 0.1900 | 0.3043 | 0.0278 |
| | (1.1466) | (1.5962) | (0.3248) |
| Dturn | -0.0294 | -0.0205 | 0.0146 |
| | (-0.5100) | (-0.3314) | (0.3494) |
| INST | -0.3703* | -0.4060* | -0.1444 |
| | (-1.7925) | (-1.7624) | (-1.1690) |
| Mshare | -0.4088 | -0.3869 | -0.1587 |
| | (-1.3207) | (-1.1376) | (-0.6893) |
| Mfee | 0.2635 | 0.2011 | 1.2142* |
| | (0.3140) | (0.2178) | (1.7127) |
| Occupy | -0.6179 | -0.8644 | -0.3125 |
| | (-0.4472) | (-0.7120) | (-0.2459) |
| Big4 | 0.0917 | 0.0795 | 0.0732 |
| | (0.9878) | (0.8402) | (0.8571) |
| _cons | -0.8757 | -0.5480 | 0.0875 |
| | (-0.8572) | (-0.4906) | (0.0942) |
| Industry | Yes | Yes | Yes |
| Year | Yes | Yes | Yes |
| N | 3633 | 3171 | 3269 |
| adj. $R^2$ | 0.084 | 0.096 | 0.112 |

*t* statistics based on robust standard error in parentheses *$p < 0.1$, **$p < 0.05$, ***$p < 0.01$

**4.3| Robustness tests**

**4.3.1 Dynamic parallel trend test**

Referring to the practices of Bertrand and mullainathan (2003) and Beck et al. (2010), we use the following intertemporal dynamic effect model to test the parallel trend of the above multi-stage double difference model to investigate whether there is a difference in the level of green technology innovation before enterprises join the list of margin trading and short selling targets.

$$LN(NGInnovation_{i,t}) = \beta_0 + \beta_1 PostList_{i,t}^{-6} + \beta_2 PostList_{i,t}^{-5} + ........ + \beta_{13} PostList_{i,t}^{+6} + \sum \beta_j \left( Control\ variables_{i,t} \right) + \sum INDUSTRYID + \sum Year + \varepsilon_{i,t}$$

In model (2), $PostList_{i,t}^{-k}$、$PostList_{i,t}^{+k}$ are dummy variables. For stocks i, k years before (after) i was transferred to the list of subject matters of margin trading and short selling $PostList_{i,t}^{-k}$ ($PostList_{i,t}^{+k}$), then the value is 1, otherwise, it is 0. We take the six years before and after i being listed in the margin trading and short selling as the observation window. In particular, $PostList_{i,t}^{-6}$ ($PostList_{i,t}^{+6}$) represent 6 years or more before (after) the representative stock i is transferred to the list of margin trading and short selling. To maintain the robustness of the research results, in the regression results of the above model, the coefficient of $PostList_{i,t}^{-k}$ should not be significant. As shown in Figure 2, we use the dynamic effect model to estimate, the dynamic change diagram of the regression coefficient of $PostList_{i,t}^{-k}$、$PostList_{i,t}^{+k}$ is made, and the solid line represents the confidence interval with the level of 95%. We found that for the relationship between margin trading, short selling and the quantity of green technology innovations, the regression coefficient of $PostList_{i,t}^{-k}$ is not significant at the 5% confidence level; but regression coefficient of $PostList_{i,t}^{+k}$ is positive at the 5% confidence level, indicating that it is around the first year that the effects of the implementation of the margin trading and short selling policy begin to appear, and there is an obvious difference in the quantity of green technology innovation between the target enterprises and non-target enterprises.

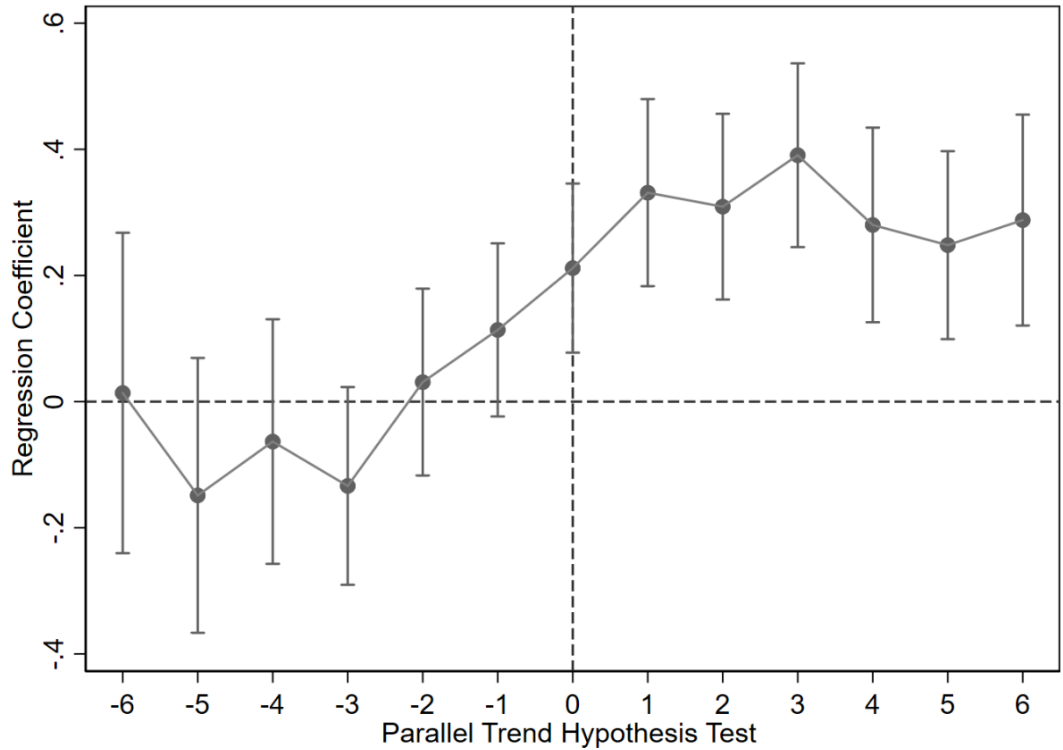

**Figure 2 Parallel trend test of margin trading, short selling and the quantity of green technology innovation**

**4.3.2 Other robustness tests**

(1) Propensity score matching. In order to overcome the selective bias between the target enterprise and the non-target enterprise, we take a series of enterprise characteristic variables and industry dummy variables described above as matching variables, use the observation data of the target enterprise and the non-target enterprise in the previous period, and use the nearest neighbor method to 1:1 match, find the non-standard enterprises with similar characteristics for the target enterprises in batches, then the matched samples were used for double difference

analysis.(2) Fixed effect model, control individual effect, and use fixed effect model for analysis.(3) In order to control the influence of regional macro factors, the fixed effect of provinces is controlled in regression analysis. (4) In order to avoid the impact of IPO on enterprises, the samples in the year of IPO are excluded. (5) China has implemented the carbon emission trading system since 2013, and has successively established carbon exchanges in Shenzhen, Beijing, Shanghai, Tianjin, Hubei, Chongqing and Guangdong provinces, in order to excluding other policy interference, the pilot policy of carbon emission trading right was controlled, and the regression results were still stable. After the above processing, the relevant results are shown in Table 5, the results of the main regression coefficients are still significantly positive, and the previous conclusions are still robust.

**Table 5-1 Robustness check**

| Inspection method | PSM | Fixed effect model | |
|---|---|---|---|
| | （1） | （2） | （3） |
| Dependent variable | $LN(NGInnovation)$ | $LN(NGInnovation)$ | $QGInnovation$ |
| *PostList* | 0.3209*** | 0.1777*** | 0.0640 |
| | (3.9500) | (2.9869) | (0.9815) |
| *List* | 0.0812 | | |
| | (0.9224) | | |
| *RD* | 4.2880*** | -3.0249 | 2.3184 |
| | (3.3255) | (-0.5864) | (1.0409) |
| *GDP* | 0.0958 | -0.1882 | 0.4114 |
| | (1.2625) | (-0.2827) | (0.9369) |
| *Lev* | 0.6300*** | -0.4925 | -0.0531 |
| | (2.8211) | (-0.8334) | (-0.2897) |
| *Growth* | 0.0673 | -0.1465 | 0.0140 |
| | (1.2908) | (-1.4552) | (0.3145) |
| Tan $qi$ | 0.4905* | -0.2515 | -0.1869 |
| | (1.7399) | (-0.2899) | (-0.6482) |
| *Cfo* | -0.5135 | -0.5608 | -0.0030 |
| | (-1.3810) | (-0.8442) | (-0.0094) |
| *Board* | -0.0600 | -0.2775 | -0.2130 |
| | (-0.3349) | (-0.7316) | (-1.4109) |
| *Indep* | 0.2922 | 0.3100 | 0.4206 |
| | (0.5246) | (0.2969) | (1.0313) |
| *Top*1 | -0.5266* | 0.6926 | 0.3490 |
| | (-1.7460) | (0.6065) | (0.9769) |
| *Soe* | 0.0773 | 0.0663 | -0.2674 |
| | (0.9211) | (0.1550) | (-1.3082) |
| *ER* | 0.2395 | 2.8145 | 2.4225* |
| | (0.4453) | (0.8526) | (1.7638) |
| *Roa* | -1.7197 | 1.8982 | -1.1710 |
| | (-0.8115) | (0.5647) | (-0.9100) |
| *Roe* | 2.7105** | 0.2129 | 0.7851 |
| | (2.1505) | (0.1298) | (1.1615) |
| *ATO* | -0.1492 | -0.1309 | -0.0710 |
| | (-1.4847) | (-0.6530) | (-0.5499) |
| *Dual* | -0.0231 | -0.0593 | -0.0470 |
| | (-0.3861) | (-0.6115) | (-0.8440) |
| *BM* | 0.0793** | -0.0172 | -0.0386** |
| | (2.2747) | (-0.3193) | (-2.1344) |
| *TobinQ* | -0.1243*** | 0.0497 | 0.0200 |
| | (-5.3482) | (0.6477) | (1.1407) |
| *Size* | 0.0002*** | -0.0001 | -0.0000 |
| | (5.0799) | (-1.2298) | (-0.4706) |
| *ListAge* | -0.0328 | 1.1626*** | 0.0722 |
| | (-0.4727) | (3.3421) | (0.5255) |
| *FirmAge* | -0.2561** | -0.8626 | 0.5225 |
| | (-2.2906) | (-1.3233) | (1.2661) |
| *Balance* | 0.0294 | 0.1603 | 0.0864 |
| | (0.4610) | (1.0786) | (0.7547) |
| *Dturn* | 0.0297 | -0.2197* | -0.0111 |
| | (0.6165) | (-1.8900) | (-0.3502) |
| *INST* | 0.6712*** | -0.2611 | -0.2675*** |
| | (4.4792) | (-1.0007) | (-2.6128) |
| *Mshare* | 0.1154 | -0.4704 | 0.1573 |
| | (0.5975) | (-0.8664) | (0.4288) |
| *Mfee* | -0.8182 | 3.7577 | -0.6745 |
| | (-1.4989) | (1.3264) | (-1.4285) |
| *Occupy* | 0.8398 | -3.4229* | -0.1091 |
| | (0.6988) | (-1.6926) | (-0.1057) |
| *Big*4 | 0.3585** | -0.2042 | -0.1137 |
| | (2.4603) | (-1.0836) | (-1.3033) |
| _cons | 0.3992 | 1.9558 | -4.7804 |
| | (0.3586) | (0.2963) | (-1.0233) |
| Industry | Yes | Yes | Yes |
| Year | Yes | Yes | Yes |
| N | 4992 | 6080 | 6080 |
| adj. $R^2$ | 0.296 | 0.077 | 0.108 |

$t$ statistics based on robust standard error in parentheses $^* p < 0.1$, $^{**} p < 0.05$, $^{***} p < 0.01$

**Table 5-2 Robustness check**

| Inspection method | Control area factors | Excluding IPO samples in the current year | Eliminate other policy interference |
|---|---|---|---|
| | （4） | （5） | （6） |
| Dependent variable | LN(NGInnovation) | LN(NGInnovation) | LN(NGInnovation) |
| PostList | 0.3542*** | 0.3449*** | 0.3489*** |
| | (4.8793) | (4.6394) | (4.6910) |
| Post2policy2 | | | 0.1295 |
| | | | (1.5641) |
| policy2 | | | 0.1033 |
| | | | (1.1162) |
| List | 0.0799 | 0.0861 | 0.0832 |
| | (0.9507) | (1.0233) | (0.9850) |
| RD | 3.7459*** | 4.0429*** | 4.0455*** |
| | (3.0594) | (3.3060) | (3.3096) |
| GDP | 0.2381 | 0.0802 | 0.0863 |
| | (0.7108) | (1.1101) | (1.1645) |
| Lev | 0.6536*** | 0.6663*** | 0.6722*** |
| | (3.2068) | (3.2375) | (3.2606) |
| Growth | 0.0736 | 0.0695 | 0.0679 |
| | (1.5683) | (1.4698) | (1.4339) |
| Tan $qi$ | 0.4129 | 0.3691 | 0.3623 |
| | (1.6429) | (1.4578) | (1.4402) |
| Cfo | -0.3356 | -0.3706 | -0.3788 |
| | (-0.9496) | (-1.0252) | (-1.0455) |
| Board | 0.0302 | 0.0239 | 0.0215 |
| | (0.1788) | (0.1414) | (0.1270) |
| Indep | 0.4088 | 0.3741 | 0.3651 |
| | (0.7905) | (0.7223) | (0.7057) |
| Top1 | -0.2281 | -0.3089 | -0.3160 |
| | (-0.8215) | (-1.0911) | (-1.1135) |
| Soe | 0.0632 | 0.0791 | 0.0797 |
| | (0.7765) | (0.9764) | (0.9738) |
| ER | 0.0656 | 0.1508 | 0.0872 |
| | (0.0446) | (0.2831) | (0.1603) |
| Roa | -0.7735 | -0.8775 | -0.8891 |
| | (-0.4189) | (-0.4650) | (-0.4711) |
| Roe | 2.1806** | 2.2386** | 2.2467** |
|  | (2.0023) | (2.0224) | (2.0297) |
| ATO | -0.0314 | -0.0405 | -0.0387 |
|  | (-0.3440) | (-0.4271) | (-0.4091) |
| Dual | 0.0059 | 0.0034 | 0.0028 |
|  | (0.1070) | (0.0608) | (0.0499) |
| BM | 0.0658* | 0.0749** | 0.0747** |
|  | (1.8460) | (2.1012) | (2.0959) |
| TobinQ | -0.1422*** | -0.1429*** | -0.1423*** |
|  | (-6.2865) | (-6.2233) | (-6.1877) |
| Size | 0.0002*** | 0.0002*** | 0.0002*** |
|  | (4.5858) | (4.7837) | (4.8143) |
| ListAge | 0.0008 | -0.0335 | -0.0366 |
|  | (0.0129) | (-0.5515) | (-0.6038) |
| FirmAge | -0.2448** | -0.2599*** | -0.2580** |
|  | (-2.4619) | (-2.5838) | (-2.5668) |
| Balance | 0.0763 | 0.0703 | 0.0699 |
|  | (1.2421) | (1.1358) | (1.1254) |
| Dturn | 0.0721* | 0.0809** | 0.0817** |
|  | (1.8416) | (2.0499) | (2.0728) |
| INST | 0.5077*** | 0.5111*** | 0.5139*** |
|  | (3.8852) | (3.9124) | (3.9325) |
| Mshare | 0.0110 | 0.0211 | 0.0179 |
|  | (0.0652) | (0.1240) | (0.1052) |
| Mfee | -0.5350 | -0.5573 | -0.5669 |
|  | (-1.0369) | (-1.0897) | (-1.1116) |
| Occupy | 0.7273 | 0.8744 | 0.8683 |
|  | (0.6304) | (0.7523) | (0.7489) |
| Big4 | 0.4887*** | 0.4564*** | 0.4477*** |
|  | (3.4561) | (3.2122) | (3.1241) |
| _cons | -1.1927 | 0.8027 | 0.6963 |
|  | (-0.3527) | (0.7586) | (0.6453) |
| Industry | Yes | Yes | Yes |
| Province | Yes |  |  |
| Year | Yes | Yes | Yes |
| N | 6080 | 6080 | 6080 |
| pseudo $R^2$ | 0.327 | 0.318 | 0.318 |

$t$ statistics based on robust standard error in parentheses *$p < 0.1$, **$p < 0.05$, ***$p < 0.01$

## 5| FURTHER RESEARCH ON MARGIN TRADING, SHORT SELLING AND THE QUANTITY OF GREEN TECHNOLOGY INNOVATION

### 5.1 Consider the performance pressure and speculative tendency of enterprises

From the motivation of enterprise green technology innovation, under the margin trading and short selling, whether the enterprise will change the investment direction due to performance pressure, whether it will change the green technology innovation strategy due to the enterprise's short-term speculation tendency, and whether above factors affects the effect of policy implementation deserves further study.

Firstly, as shown in columns (1) - (2) of table 6, this paper divides the whole sample into two groups according to whether the earnings per share decreased($EPS\_down$) compared with the previous period. In addition, as shown in columns (3) - (4) of table 6, we calculated the gap between the return on investment of financial assets and the return on investment of operating assets($Gap$). According to the upper quartile and lower quartile of the return gap, we took out the two groups with high and low return differences for regression analysis respectively.

The regression results show that when facing the pressure of performance decline, the promotion effect of policy implementation decreases. We believe that when facing the pressure of performance, enterprises will reduce innovation expenditure, and their tendency of engaging green innovation is relatively low. In the group with a high gap between financial assets and operating assets, we found that the promotion effect of policy was relatively low. We think it is due to the high income difference between financial assets and operating assets, enterprises may participate in short-term financial speculation, which means that the behavior mode of such enterprises is short-term and their tendency of engaging green innovation is relatively low.

**5.2 Consider the risk of stock prices crash**

In order to directly investigate whether there is a significant difference in the promotion of policy innovation when enterprises face stock price pressure under the margin trading and short selling, this paper uses the fluctuation ratio of yield ($DUVOL$) to reflect the downward trend and collapse risk of individual stock prices, and divides the samples into two groups according to the annual median. The results of grouped regression are reported in columns (5) - (6) of table 6. We find that the promotion effect of policies is relatively higher and in the groups with high risk of stock price collapse. We believe that companies with high risk of share price collapse face higher pressure of short selling, so the short selling mechanism plays a stronger role in driving the quantity of green technology innovation.

Table 6-1 Futher Study

| Topic | Declining performance | Performance did not decline | High yield gap |
|---|---|---|---|
| | （1） | （2） | （3） |
| Dependent variable | $LN(NGInnovation)$ | $LN(NGInnovation)$ | $LN(NGInnovation)$ |
| $PostList$ | 0.2744*** | 0.4268*** | 0.3765** |
| | (2.8806) | (4.6647) | (2.5200) |
| $List$ | 0.1580 | -0.0009 | -0.0819 |
| | (1.4885) | (-0.0091) | (-0.5207) |
| $RD$ | 4.0137** | 4.0047*** | 5.0642** |
| | (2.5035) | (3.0666) | (2.2705) |
| $GDP$ | 0.0411 | 0.1090 | 0.0590 |
| | (0.5242) | (1.3068) | (0.4860) |
| $Lev$ | 0.4647* | 0.8226*** | 0.8994** |
| | (1.7392) | (3.1377) | (2.3162) |
| $Growth$ | 0.0240 | 0.1208** | 0.0612 |
| | (0.2572) | (2.1313) | (0.6796) |
| $\text{Tan}\,qi$ | 0.6655** | 0.1182 | 0.8774** |
| | (2.1728) | (0.3860) | (2.0551) |
| $Cfo$ | -0.1951 | -0.2804 | -0.1795 |
| | (-0.3927) | (-0.6263) | (-0.2681) |
| $Board$ | 0.0581 | -0.0062 | 0.0207 |
| | (0.3018) | (-0.0335) | (0.0807) |
| $Indep$ | 0.0987 | 0.5685 | 0.3345 |
| | (0.1649) | (0.9745) | (0.3618) |
| Top1 | -0.2157 | -0.4482 | -0.1000 |
| | (-0.6533) | (-1.3271) | (-0.2080) |
| Soe | 0.0516 | 0.0933 | -0.1160 |
| | (0.5746) | (1.0307) | (-0.9379) |
| ER | 0.5941 | 0.2372 | 0.1938 |
| | (1.0719) | (0.3524) | (0.1904) |
| Roa | -4.4805 | 1.2140 | -0.2377 |
| | (-1.6048) | (0.5677) | (-0.0755) |
| Roe | 4.4069** | 1.3080 | 1.4565 |
| | (2.5614) | (1.0711) | (0.7936) |
| ATO | -0.0672 | -0.0194 | 0.1336 |
| | (-0.5861) | (-0.1872) | (0.9495) |
| Dual | -0.0045 | 0.0179 | 0.0283 |
| | (-0.0677) | (0.2749) | (0.3028) |
| BM | 0.1206** | 0.0514 | 0.1925*** |
| | (2.5291) | (1.3786) | (3.5950) |
| TobinQ | -0.1104*** | -0.1692*** | -0.1185** |
| | (-3.6931) | (-6.1481) | (-2.5649) |
| Size | 0.0002*** | 0.0002*** | 0.0002*** |
| | (4.4260) | (4.0868) | (4.3387) |
| ListAge | -0.0756 | 0.0211 | -0.0904 |
| | (-1.1043) | (0.2948) | (-0.8889) |
| FirmAge | -0.1683 | -0.3689*** | -0.1525 |
| | (-1.5021) | (-3.2724) | (-0.9626) |
| Balance | 0.0737 | 0.0565 | 0.1731 |
| | (0.9850) | (0.7788) | (1.5603) |
| Dturn | 0.1023* | 0.0583 | 0.0453 |
| | (1.9253) | (0.9907) | (0.4464) |
| INST | 0.4343*** | 0.5782*** | 0.4427* |
| | (2.6651) | (3.5661) | (1.9309) |
| Mshare | 0.0766 | -0.0672 | 0.1382 |
| | (0.3803) | (-0.3283) | (0.4257) |
| Mfee | -0.2629 | -0.8313 | -1.2345 |
| | (-0.4241) | (-1.4135) | (-1.4137) |
| Occupy | 1.1863 | 0.9117 | -1.7845 |
| | (0.8205) | (0.6118) | (-0.9306) |
| Big4 | 0.5209*** | 0.4040*** | 0.5342** |
| | (3.3010) | (2.6686) | (2.4906) |
| _cons | 0.1982 | 1.2446 | -0.0493 |
| | (0.1637) | (1.0158) | (-0.0293) |
| Industry | Yes | Yes | Yes |
| Year | Yes | Yes | Yes |
| $N$ | 2793 | 3287 | 1221 |
| adj. $R^2$ | 0.305 | 0.332 | 0.336 |

**Table 6-2 Futher Study**

| Topic | Low yield gap | High risk of collapse | Low risk of collapse |
|---|---|---|---|
| | （4） | （5） | （6） |
| Dependent variable | $LN(NGInnovation)$ | $LN(NGInnovation)$ | $LN(NGInnovation)$ |
| *PostList* | 0.4802*** | 0.4204*** | 0.2499*** |
| | (3.2727) | (4.4034) | (2.7235) |
| *List* | 0.1260 | -0.0305 | 0.2122** |
| | (0.8447) | (-0.3158) | (2.0630) |
| *RD* | 3.0766 | 4.3611*** | 3.9272** |
| | (1.3085) | (3.0532) | (2.4984) |
| *GDP* | 0.1085 | 0.0332 | 0.1293 |
| | (0.7630) | (0.4297) | (1.4782) |
| *Lev* | 0.6176 | 0.7848*** | 0.5980** |
| | (1.4031) | (3.2188) | (2.2903) |
| *Growth* | 0.1007 | 0.0310 | 0.0872 |
| | (1.0811) | (0.5097) | (1.2548) |
| Tan *qi* | 0.5120 | 0.3564 | 0.4533 |
| | (0.9839) | (1.2817) | (1.2860) |
| *Cfo* | 0.4484 | -0.4639 | -0.1851 |
| | (0.6464) | (-1.0577) | (-0.3744) |
| *Board* | -0.1880 | -0.0537 | 0.0749 |
| | (-0.6528) | (-0.3028) | (0.3492) |
| *Indep* | -0.2502 | 0.5966 | 0.1006 |
| | (-0.2844) | (1.0493) | (0.1476) |
| *Top*1 | -1.3007*** | -0.5520* | -0.0841 |
| | (-2.7096) | (-1.7049) | (-0.2362) |
| *Soe* | 0.2412* | 0.1352 | 0.0241 |
| | (1.7168) | (1.4847) | (0.2581) |
| *ER* | 0.2382 | 0.1960 | 0.0799 |
| | (0.1891) | (0.3502) | (0.1228) |
| *Roa* | -0.8141 | -0.1169 | -1.4647 |
| | (-0.2358) | (-0.0532) | (-0.6059) |
| *Roe* | 2.5190 | 2.6977** | 1.8030 |
| | (1.2842) | (2.2126) | (1.2619) |
| *ATO* | -0.1372 | -0.1033 | -0.0104 |
| | (-0.7806) | (-0.9557) | (-0.0907) |
| *Dual* | -0.0217 | -0.0317 | 0.0448 |
| | (-0.2055) | (-0.5098) | (0.6400) |
| *BM* | 0.0467 | 0.0568 | 0.0942** |
| | (0.8906) | (1.5640) | (2.0869) |
| TobinQ | -0.1534*** | -0.1694*** | -0.1224*** |
| | (-4.3131) | (-6.5525) | (-3.8559) |
| Size | 0.0003** | 0.0003*** | 0.0002*** |
| | (2.5416) | (4.6041) | (4.1822) |
| ListAge | -0.1687 | -0.0740 | 0.0129 |
| | (-1.5202) | (-1.1001) | (0.1669) |
| FirmAge | -0.1765 | -0.1314 | -0.4106*** |
| | (-1.0158) | (-1.2433) | (-3.2085) |
| Balance | -0.1111 | 0.0455 | 0.0839 |
| | (-1.0527) | (0.6960) | (0.9982) |
| Dturn | 0.1396 | 0.1373** | 0.0370 |
| | (1.4693) | (2.3836) | (0.6525) |
| INST | 0.6465*** | 0.4537*** | 0.5408*** |
| | (2.6164) | (2.7180) | (3.2674) |
| Mshare | 0.1751 | 0.0487 | -0.0121 |
| | (0.5511) | (0.2459) | (-0.0554) |
| Mfee | -1.5292 | -0.5252 | -0.4979 |
| | (-1.1921) | (-0.8926) | (-0.7441) |
| Occupy | 4.8800** | -0.2155 | 1.9199 |
| | (1.9834) | (-0.1606) | (1.3241) |
| Big4 | 0.1623 | 0.4211** | 0.4619*** |
| | (0.8962) | (2.4818) | (3.0035) |
| _cons | 1.0615 | 1.4550 | 0.2000 |
| | (0.5955) | (1.2462) | (0.1541) |
| Industry | Yes | Yes | Yes |
| Year | Yes | Yes | Yes |
| N | 1229 | 3053 | 3025 |
| adj. $R^2$ | 0.329 | 0.315 | 0.323 |

$t$ statistics based on robust standard error in parentheses * $p < 0.1$, ** $p < 0.05$, *** $p < 0.01$

**5.3 Consider management shareholding and institutional investor shareholding**

We conduct grouping regression on the samples according to the management shareholding ratio( $Mshare$ ) and institutional shareholding ratio( $INST$ ). As shown in columns (1) - (4) of table 7, this paper examines the impact of margin trading and short selling on the quantity of green innovation of enterprise when the shareholding level of enterprise management is different from that of institutional investors. We found that the policy promotion effect is not significant in the group with high shareholding ratio of management. We believe that the reason is that when the management holds the company's shares, especially when the shareholding level is high, the enterprise is relatively weak under the supervision of external stakeholders, so the tendency of relevant enterprises to engage in dual externalities such as green innovation is relatively low. In addition, we found that in the group with high institutional shareholding ratio, the role of policy promotion is stronger. We believe that institutional investors, as the representative of external stakeholders, play a supervisory role, and relevant enterprises have stronger motivation to engage in green technology innovation.

**5.4 Consider product market competition**

The stronger the competition in the industry where the enterprise is located, in order to prevent being eliminated, more resources will be used for business activities and industrial investment to improve the core competitiveness of the enterprise. The weak product market competition creates a comfortable living environment for enterprises. The competitive environment pressure faced by enterprises will affect the tendency of engaging green technology innovation. In order to explore its impact on the policy effect, this paper uses the Herfindal index, that is, the sum of the square of the proportion of the operating revenue of Companies to operating revenue of all companies in the industry, to measure the product market competition. It is divided into two groups according to the annual median of Herfindal index( *HHI* ). If it is lower than this median, the industry competition is strong; Otherwise, the industry competition is weak. Columns (5) and (6) of Table 7 report the regression results under the market competition of different products. We found that the promotion effect of policies is positive significance in the group with low competitive pressure, and it was not significant in the high competitive pressure group. Combined with the following research（Section 6）, we believe that under the high-intensity competition and facing the limited resources, enterprises will adopt the green development strategy of paying attention to quality rather than quantity.

Table 7-1 Futher Study

| Topic | Management shareholding ratio grouping (high) | Management shareholding ratio grouping (low) | institutional shareholding ratio (high) | Institutional shareholding ratio grouping (low) |
|---|---|---|---|---|
| | （1） | （2） | （3） | （4） |
| Dependent variable | $LN(NGInnovation)$ | $LN(NGInnovation)$ | $LN(NGInnovation)$ | $LN(NGInnovation)$ |
| PostList | 0.1373 | 0.3664*** | 0.3535*** | 0.2648*** |
| | (1.2289) | (3.8204) | (3.3862) | (2.7477) |
| List | 0.1048 | 0.0803 | 0.0966 | 0.0879 |
| | (0.8858) | (0.7353) | (0.8244) | (0.8437) |
| RD | 4.3318*** | 5.1675** | 4.7910** | 3.9379*** |
| | (3.6889) | (2.1517) | (2.1611) | (3.3772) |
| GDP | 0.0179 | 0.1195 | 0.1168 | 0.0407 |
| | (0.2153) | (1.1216) | (1.0133) | (0.5128) |
| Lev | 0.6898** | 0.8928*** | 1.0143*** | 0.3659 |
| | (2.5461) | (2.9926) | (3.1767) | (1.4617) |
| Growth | 0.0148 | 0.1261* | 0.1218 | 0.0266 |
| | (0.2358) | (1.8479) | (1.6412) | (0.4546) |
| $\text{Tan}\,qi$ | 0.5887** | 0.3850 | 0.3554 | 0.4278 |
| | (2.0957) | (0.8785) | (0.8447) | (1.5235) |
| Cfo | 0.3604 | -0.7265 | -1.0535* | 0.1140 |
| | (0.8244) | (-1.3418) | (-1.8789) | (0.2804) |
| Board | -0.0460 | -0.0145 | 0.2246 | -0.2328 |
| | (-0.2388) | (-0.0619) | (0.8782) | (-1.3452) |
| Indep | 0.0024 | 0.7813 | 0.9201 | -0.4544 |
| | (0.0035) | (1.1474) | (1.2434) | (-0.7725) |
| Top1 | -0.8312** | 0.0985 | 0.0639 | -0.6699** |
| | (-2.1562) | (0.2489) | (0.1307) | (-2.1479) |
| Soe | 0.1288 | 0.0912 | 0.0120 | 0.1698* |
| | (1.0594) | (0.8723) | (0.1124) | (1.7730) |
| ER | 0.0410 | 0.4028 | 0.0030 | 0.3832 |
| | (0.0677) | (0.4672) | (0.0029) | (0.6741) |
| Roa | -2.2604 | 3.5404 | 1.6458 | -2.9766 |
| | (-0.9322) | (1.3735) | (0.6638) | (-1.0886) |
| Roe | 2.6019* | 0.2143 | 1.3164 | 2.9350* |
| | (1.7306) | (0.1610) | (0.9606) | (1.7972) |
| ATO | -0.3415*** | 0.0207 | -0.1277 | 0.0289 |
| | (-2.7068) | (0.1669) | (-0.9292) | (0.2684) |
| Dual | 0.0554 | -0.1401* | 0.0071 | 0.0015 |

|  | (0.8197) | (-1.6697) | (0.0755) | (0.0254) |
|---|---|---|---|---|
| BM | 0.0113 | 0.0665 | 0.0054 | 0.1255** |
|  | (0.1872) | (1.6268) | (0.1272) | (2.3894) |
| TobinQ | -0.1544*** | -0.1297*** | -0.1478*** | -0.1382*** |
|  | (-5.9783) | (-3.3268) | (-4.3205) | (-5.0405) |
| Size | 0.0016*** | 0.0002*** | 0.0002*** | 0.0006** |
|  | (9.5203) | (4.7743) | (4.7559) | (2.4335) |
| ListAge | -0.1330* | 0.1244 | 0.0151 | -0.0474 |
|  | (-1.9494) | (1.3995) | (0.1478) | (-0.7651) |
| FirmAge | -0.1248 | -0.3819** | -0.4248** | -0.1156 |
|  | (-1.1744) | (-2.0329) | (-2.4970) | (-1.1742) |
| Balance | 0.0213 | 0.0608 | 0.1074 | 0.0454 |
|  | (0.3204) | (0.5355) | (0.8913) | (0.7315) |
| Dturn | 0.0877* | 0.0542 | 0.1840** | 0.0279 |
|  | (1.8795) | (0.8750) | (2.2964) | (0.6175) |
| INST | 0.5796*** | 0.4151** | 0.1407 | 0.5603*** |
|  | (3.3964) | (2.2052) | (0.3788) | (2.6549) |
| Mshare | -0.0925 | 5.1366 | 0.0020 | 0.2836 |
|  | (-0.4792) | (0.3355) | (0.0053) | (1.6127) |
| Mfee | -0.5641 | -0.8663 | -1.1547 | -0.1926 |
|  | (-1.0307) | (-0.8417) | (-1.1476) | (-0.3940) |
| Occupy | -1.1891 | 1.0841 | 2.1843 | 0.1402 |
|  | (-0.7306) | (0.7546) | (1.3169) | (0.0985) |
| Big4 | 0.3316 | 0.3442** | 0.3900** | 0.6246*** |
|  | (1.6001) | (2.2106) | (2.4635) | (3.0476) |
| _cons | 1.6121 | 0.2577 | -0.9067 | 2.3902** |
|  | (1.3224) | (0.1759) | (-0.5846) | (2.0923) |
| Industry | Yes | Yes | Yes | Yes |
| Year | Yes | Yes | Yes | Yes |
| N | 3054 | 3026 | 3003 | 3077 |
| adj. $R^2$ | 0.311 | 0.366 | 0.346 | 0.267 |

$t$ statistics based on robust standard error in parentheses * $p < 0.1$, ** $p < 0.05$, *** $p < 0.01$

**Table 7-2 Futher Study**

| Topic | Industry competitiveness grouping (high) | Industry competitiveness grouping (low) |
|---|---|---|
|  | （5） | （6） |

| Dependent variable | $LN(NGInnovation)$ | $LN(NGInnovation)$ |
|---|---|---|
| PostList | 0.1685 | 0.4176*** |
|  | (1.5953) | (4.1845) |
| List | 0.0644 | 0.0955 |
|  | (0.5201) | (0.8636) |
| RD | 0.8694 | 5.9835*** |
|  | (0.4547) | (3.6046) |
| GDP | -0.0293 | 0.1421 |
|  | (-0.2552) | (1.5025) |
| Lev | 0.4585 | 0.7584*** |
|  | (1.5888) | (2.6188) |
| Growth | 0.0129 | 0.0873 |
|  | (0.1755) | (1.4588) |
| Tan$qi$ | 0.0054 | 0.5176 |
|  | (0.0159) | (1.4969) |
| Cfo | -0.0714 | -0.6934 |
|  | (-0.1356) | (-1.4388) |
| Board | -0.2081 | 0.1484 |
|  | (-0.9181) | (0.6234) |
| Indep | 0.3191 | 0.1184 |
|  | (0.4716) | (0.1595) |
| Top1 | -0.6575 | -0.0005 |
|  | (-1.5797) | (-0.0013) |
| Soe | -0.0431 | 0.1750* |
|  | (-0.3748) | (1.6520) |
| ER | 1.4648 | 0.4591 |
|  | (1.2918) | (0.7787) |
| Roa | 0.9160 | -1.4891 |
|  | (0.3434) | (-0.5851) |
| Roe | 0.8631 | 2.7695* |
|  | (0.5478) | (1.8889) |
| ATO | 0.1321 | -0.1071 |
|  | (0.8673) | (-0.9265) |
| Dual | -0.0785 | 0.0648 |
|  | (-1.0235) | (0.8542) |
| BM | 0.0335 | 0.0383 |
|  | (0.6255) | (0.9238) |
| TobinQ | -0.1479*** | -0.1481*** |
|  | (-4.5943) | (-4.8278) |
| Size | 0.0008*** | 0.0002*** |
|  | (6.1279) | (4.0318) |
| ListAge | 0.0912 | -0.1277 |
|  | (1.0840) | (-1.5681) |
| *FirmAge* | -0.1659 | -0.2608** |
| | (-1.1693) | (-2.0175) |
| *Balance* | 0.0893 | 0.0389 |
| | (1.0157) | (0.4874) |
| *Dturn* | 0.0373 | 0.0983* |
| | (0.6810) | (1.7908) |
| *INST* | 0.5180*** | 0.4514*** |
| | (2.6039) | (2.7407) |
| *Mshare* | 0.0720 | -0.0684 |
| | (0.3160) | (-0.2856) |
| *Mfee* | -0.4437 | -0.4684 |
| | (-0.6035) | (-0.6617) |
| *Occupy* | 0.6428 | 1.6216 |
| | (0.3923) | (1.0356) |
| *Big*4 | 0.1584 | 0.5962*** |
| | (0.8529) | (3.2808) |
| _cons | 2.7621* | -0.0988 |
| | (1.8557) | (-0.0692) |
| Industry | Yes | Yes |
| Year | Yes | Yes |
| N | 2685 | 3395 |
| adj. $R^2$ | 0.310 | 0.345 |

*t* statistics based on robust standard error in parentheses* $p < 0.1$, ** $p < 0.05$, *** $p < 0.01$

**5.5 Consider the differences between margin trading and short selling**

In order to distinguish the differential impact of margin trading and short selling, we use the proportion of margin trading balance to circulation market value( *Magrin* ) and the proportion of short selling balance to circulation market value( *Short* ) as proxy variables of margin trading intensity and short selling intensity respectively. We used the above variables annual median as the grouping basis for grouping regression, according to the regression results in columns (1) - (2) of table 8, in the high short selling intensity group, the promotion effect of policy is significant at the 1% confidence level, and in the low short selling intensity group, the promotion effect of policy is not significant. According to the regression results in columns (3) - (4) of table 8, in the high margin trading intensity group, the promotion effect of the policy is relatively lower than that in the low margin trading intensity group. The results show that short selling mechanism is an important reason to promote enterprises to engage in the quantity of green technology innovation, while margin trading has negative effects.

**Table 8 Futher Study**

| Topic | Short selling (High intensity) | Short selling (Low intensity) | Margin tading (High intensity) | Margin tading (Low intensity) |
|---|---|---|---|---|
| | （1） | （2） | （3） | （4） |
| Dependent variable | $LN(NGInnovation)$ | $LN(NGInnovation)$ | $LN(NGInnovation)$ | $LN(NGInnovation)$ |
| PostList | 0.3275*** | -0.0561 | 0.2923*** | 0.5432*** |
| | (3.1115) | (-0.5284) | (3.2485) | (2.8391) |
| List | -0.2039 | 0.0606 | 0.1086 | 0.0447 |
| | (-1.1165) | (0.6423) | (0.6821) | (0.4498) |
| RD | 5.2421*** | 2.4031* | 5.8782*** | 2.0114 |
| | (2.6365) | (1.7702) | (3.1740) | (1.4689) |
| GDP | 0.2427* | -0.0307 | 0.2295* | -0.0379 |
| | (1.8937) | (-0.3615) | (1.8731) | (-0.4294) |
| Lev | 0.6962* | 0.5805*** | 0.7412** | 0.6099*** |
| | (1.7852) | (2.7223) | (2.0474) | (2.8001) |
| Growth | 0.0862 | 0.0628 | 0.0796 | 0.0646 |
| | (0.9597) | (1.1915) | (0.9414) | (1.1852) |
| Tan $qi$ | 0.5632 | 0.5757** | 0.2410 | 0.6624** |
| | (1.1192) | (2.2091) | (0.5225) | (2.4642) |
| Cfo | -1.2457* | 0.3157 | -1.1280* | 0.3211 |
| | (-1.7918) | (0.8606) | (-1.8300) | (0.8293) |
| Board | 0.2200 | -0.1389 | 0.1540 | -0.0464 |
| | (0.8101) | (-0.8728) | (0.5934) | (-0.2831) |
| Indep | 1.0641 | -0.0005 | 0.9672 | 0.1090 |
| | (1.4321) | (-0.0009) | (1.3266) | (0.1848) |
| Top1 | -0.2211 | -0.2077 | -0.0704 | -0.3285 |
| | (-0.4388) | (-0.6822) | (-0.1477) | (-1.0760) |
| Soe | 0.1462 | -0.0344 | 0.1299 | -0.0185 |
| | (1.1460) | (-0.4252) | (1.0542) | (-0.2234) |
| ER | 1.2102* | 1.1566 | 0.9990 | 1.0982 |
| | (1.7517) | (1.4013) | (1.4747) | (1.2676) |
| Roa | -1.7733 | 0.0262 | -1.4746 | -0.4971 |
| | (-0.5330) | (0.0142) | (-0.4614) | (-0.2636) |
| Roe | 3.3294* | 0.9724 | 3.2174* | 1.3375 |
| | (1.7449) | (0.9507) | (1.7584) | (1.2834) |
| ATO | -0.1073 | 0.0482 | -0.1075 | 0.0313 |

|          | (-0.7386)   | (0.5025)    | (-0.7839)   | (0.3144)    |
|----------|-------------|-------------|-------------|-------------|
| Dual     | -0.0524     | 0.0229      | -0.0775     | 0.0379      |
|          | (-0.5236)   | (0.3998)    | (-0.8212)   | (0.6267)    |
| BM       | 0.0447      | 0.1105***   | 0.0433      | 0.1264***   |
|          | (0.9545)    | (2.9592)    | (0.9377)    | (3.1141)    |
| TobinQ   | -0.1529***  | -0.1416***  | -0.1443***  | -0.1414***  |
|          | (-3.8391)   | (-5.4250)   | (-3.7749)   | (-5.3909)   |
| Size     | 0.0002***   | 0.0004**    | 0.0002***   | 0.0002***   |
|          | (4.5843)    | (2.3712)    | (4.3924)    | (3.3393)    |
| ListAge  | -0.0103     | -0.0265     | -0.0109     | -0.0391     |
|          | (-0.0728)   | (-0.4733)   | (-0.0811)   | (-0.6962)   |
| FirmAge  | -0.4156*    | -0.2151**   | -0.3239     | -0.2727***  |
|          | (-1.8692)   | (-2.1502)   | (-1.5978)   | (-2.7076)   |
| Balance  | 0.1665      | 0.0593      | 0.1698      | 0.0399      |
|          | (1.3640)    | (1.0312)    | (1.5124)    | (0.6650)    |
| Dturn    | 0.0657      | 0.0715*     | 0.0982      | 0.0761*     |
|          | (0.6519)    | (1.7435)    | (1.1143)    | (1.8264)    |
| INST     | 0.1772      | 0.6250***   | 0.1664      | 0.6517***   |
|          | (0.6830)    | (4.4495)    | (0.6995)    | (4.4460)    |
| Mshare   | -0.3518     | 0.1362      | -0.1778     | 0.1189      |
|          | (-0.7454)   | (0.8107)    | (-0.4075)   | (0.6970)    |
| Mfee     | -1.0234     | 0.0797      | -1.4155*    | 0.1703      |
|          | (-1.1546)   | (0.1388)    | (-1.6630)   | (0.2864)    |
| Occupy   | 0.2315      | 1.6978      | -0.0262     | 1.0817      |
|          | (0.1077)    | (1.4271)    | (-0.0143)   | (0.8156)    |
| Big4     | 0.4159***   | 0.2981*     | 0.4461***   | 0.3331**    |
|          | (2.5884)    | (1.8647)    | (2.7705)    | (2.1498)    |
| _cons    | -2.0230     | 2.2668**    | -1.9462     | 2.3896**    |
|          | (-1.0643)   | (1.9905)    | (-1.1053)   | (1.9931)    |
| Industry | Yes         | Yes         | Yes         | Yes         |
| Year     | Yes         | Yes         | Yes         | Yes         |
| N        | 2355        | 3725        | 2577        | 3503        |
| adj. $R^2$ | 0.388     | 0.204       | 0.384       | 0.218       |

$t$ statistics based on robust standard error in parentheses $^* p < 0.1$, $^{**} p < 0.05$, $^{***} p < 0.01$

### 5.6 Considering the strength of formal environmental regulation

According to the upper quartile and lower quartile of formal environmental regulation intensity, the samples are divided into high-intensity formal environmental regulation group and low-intensity formal environmental regulation group. According to the results in Table 9 (1) - (2), we find that the promotion effect of margin trading and short selling policy on the quantity of green technology innovation is only significant in the low-intensity formal environmental regulation group. We believe that the reason for this phenomenon is that the short selling threat brought by the margin trading and short selling mechanism is more similar to an informal system spontaneously formed by the market. It plays a similar role as a public voluntary environmental regulation tool and is essentially a supplement to the government's environmental regulation. Therefore, when the formal environmental regulation is strong, the supplementary effect of margin trading and short selling policy is relatively low, but when the formal environmental regulation is weak, the supplementary effect of margin trading and short selling policy is strong.

**Table 9 Futher Study**

| Topic | Strength of environmental regulation (high) | Strength of environmental regulation (low) |
|---|---|---|
| | （1） | （2） |
| Dependent variable | $LN(NGInnovation)$ | $LN(NGInnovation)$ |
| PostList | 0.1880 | 0.4280*** |
| | (1.2579) | (3.0833) |
| List | 0.2024 | -0.0520 |
| | (1.1916) | (-0.3713) |
| RD | 5.0843 | 3.7409** |
| | (1.6139) | (2.2614) |
| GDP | 0.2130* | 0.1902 |
| | (1.6663) | (1.2864) |
| Lev | 0.6048 | 0.8695*** |
| | (1.4668) | (2.7292) |
| Growth | 0.0694 | -0.0104 |
| | (0.7605) | (-0.1202) |
| Tan $qi$ | 0.6952 | 0.7788* |
|         | (1.2507)   | (1.7343)   |
| *Cfo*   | 0.6819     | -0.3923    |
|         | (0.9767)   | (-0.6308)  |
| *Board* | 0.1667     | -0.0324    |
|         | (0.5973)   | (-0.0912)  |
| *Indep* | 2.8355***  | -0.8983    |
|         | (2.8673)   | (-0.9352)  |
| *Top*1  | 1.0926**   | -0.9155**  |
|         | (2.2017)   | (-1.9732)  |
| *Soe*   | -0.1410    | 0.1037     |
|         | (-1.0657)  | (0.7990)   |
| *ER*    | 0.4073     | 4.3090     |
|         | (0.6792)   | (1.4313)   |
| *Roa*   | 0.5869     | -1.3135    |
|         | (0.1836)   | (-0.4727)  |
| *Roe*   | 0.5657     | 2.7682*    |
|         | (0.3144)   | (1.6949)   |
| *ATO*   | -0.1847    | -0.1511    |
|         | (-1.2494)  | (-0.9518)  |
| *Dual*  | -0.1776    | -0.0116    |
|         | (-1.6331)  | (-0.1408)  |
| *BM*    | 0.0776     | 0.1036*    |
|         | (1.3118)   | (1.6484)   |
| *TobinQ*| -0.1689*** | -0.1586*** |
|         | (-3.4429)  | (-4.2689)  |
| *Size*  | 0.0002***  | 0.0002     |
|         | (3.7298)   | (1.2801)   |
| *ListAge* | -0.1172  | -0.0369    |
|         | (-0.9546)  | (-0.3970)  |
| *FirmAge* | -0.2025  | -0.1506    |
|         | (-0.8882)  | (-1.0815)  |
| *Balance* | 0.2477*  | -0.0104    |
|         | (1.7398)   | (-0.1157)  |
| *Dturn* | 0.1481     | 0.1083     |
|         | (1.5014)   | (1.5298)   |
| *INST*  | 0.1676     | 0.8020***  |
|         | (0.6367)   | (3.7486)   |
| *Mshare*| -0.0982    | 0.0558     |
|         | (-0.2123)  | (0.2262)   |
| *Mfee*  | -1.3337    | -0.3648    |
|         | (-1.5683)  | (-0.3609)  |
| *Occupy*| -0.1346    | -2.7068    |
|         | (-0.0583)  | (-1.5504)  |
| *Big* 4 | 0.3762     | 0.2977     |

|          | (1.5477)  | (1.2640) |
|----------|-----------|----------|
| _cons    | -0.6978   | 0.2224   |
|          | (-0.3773) | (0.1047) |
| Industry | Yes       | Yes      |
| Year     | Yes       | Yes      |
| $N$      | 1127      | 1937     |
| adj. $R^2$ | 0.357   | 0.284    |

$t$ statistics based on robust standard error in parentheses $^{*}p < 0.1$, $^{**}p < 0.05$, $^{***}p < 0.01$

# 6| EXPLORE THE REASONS WHY MARGIN TRADING AND SHORT SELLING CAN NOT PROMOTE THE QUALITY OF GREEN TECHNOLOGY INNOVATION

This study found that the policy did not significantly promote the quality of green technology innovation. The possible reason we first considered was the agency problem. We divided the samples into high and low groups according to the shareholding ratio of management, according to the results in columns (1) - (2) in table 10, we find that the effect of policies on the improvement of green technology innovation quality is significant in the group with low shareholding ratio of management, indicating that external governance obstacles are one of the important reasons that hinder policies from promoting green technology innovation quality, and reasonable long-term oriented external corporate governance is a key issue for the improvement of long-term patent quality.

In addition, we perform sub-sample regression according to the degree of industry competition. According to the columns (3)-(4) in Table 10, we find that under a high-intensity external competition environment, companies will take actions to improve the quality of green technological innovation rather than quantity (In conjunction with Section 5.4), it is explained that an important reason for the ineffectiveness of the policy is the insufficient promotion of other external forces to the improvement of the quality of corporate green innovation. For example, there may be a lack of long-term oriented government regulations and imperfect supervision frameworks, so that companies have sufficient motivation to improve the quality of green technological innovation only in an environment facing competitive pressures.

Table 10  Futher Study

| Topic | Management shareholding ratio grouping (high) | Management shareholding ratio grouping (low) | Industry competitiveness grouping (high) | Industry competitiveness grouping (low) |
|---|---|---|---|---|
| | （1） | （2） | （3） | （4） |
| Dependent variable | QGInnovation | QGInnovation | QGInnovation | QGInnovation |
| PostList | 0.1078 | 0.1205** | 0.2733*** | 0.0248 |
| | (0.8188) | (1.9743) | (3.0329) | (0.2380) |
| List | -0.0060 | -0.0896 | -0.0785 | -0.0317 |
| | (-0.0394) | (-0.9898) | (-0.8079) | (-0.2371) |
| RD | 4.4222** | 9.3115* | 7.0020* | 3.5581* |
| | (2.0802) | (1.6859) | (1.6769) | (1.6616) |
| GDP | -0.0350 | 0.0254 | 0.1276 | -0.0454 |
| | (-0.2610) | (0.2276) | (1.2244) | (-0.4625) |
| Lev | -0.6838** | 0.2022 | 0.3704 | -0.6927** |
| | (-2.1875) | (0.3433) | (0.6353) | (-2.1288) |
| Growth | 0.1546 | -0.0140 | 0.0118 | 0.0955 |
| | (1.6326) | (-0.2208) | (0.2190) | (0.9308) |
| $Tan_{qi}$ | 0.4884 | 0.5180 | 0.5413 | 0.2832 |
| | (1.1636) | (0.9569) | (1.3649) | (0.5943) |
| Cfo | 0.7147 | -0.2068 | -0.5430 | 0.7472 |
| | (1.3754) | (-0.3816) | (-1.3611) | (1.1469) |
| Board | -0.3450* | 0.0975 | -0.0989 | -0.0996 |
| | (-1.8060) | (0.6173) | (-0.5809) | (-0.4932) |
| Indep | -1.3630** | 0.4651 | -0.9048 | 0.1374 |
| | (-2.0738) | (0.9542) | (-1.5204) | (0.1901) |
| $Top1$ | 1.5472** | -0.5291 | 0.4722 | 0.2663 |
| | (1.9802) | (-1.1428) | (0.9379) | (0.4009) |
| Soe | 0.0458 | -0.0455 | -0.1145 | -0.0388 |
| | (0.4169) | (-0.4703) | (-1.1852) | (-0.3488) |
| ER | 2.2106 | 0.1786 | 0.4302 | 1.1417 |
| | (0.9780) | (0.2152) | (0.5220) | (1.1489) |
| Roa | -4.8996* | 5.7738 | 7.8271 | -5.6371** |
| | (-1.7802) | (0.7468) | (1.0168) | (-2.3735) |
| Roe | 1.8321 | -1.9983 | -4.1329 | 2.4174** |
|         | (1.1681)  | (-0.6017) | (-1.0506) | (2.2352)  |
| ATO     | -0.3964*  | -0.2754*  | -0.2264*  | -0.2364   |
|         | (-1.6661) | (-1.7223) | (-1.8980) | (-1.2962) |
| Dual    | -0.0699   | -0.0570   | -0.0134   | -0.0925   |
|         | (-0.8607) | (-0.4292) | (-0.1380) | (-0.9869) |
| BM      | -0.0401   | 0.0359    | 0.0237    | -0.0458*  |
|         | (-0.8394) | (0.7990)  | (0.5395)  | (-1.7208) |
| TobinQ  | 0.0301    | 0.1733    | 0.1784*   | -0.0070   |
|         | (1.0088)  | (1.4140)  | (1.7040)  | (-0.2463) |
| Size    | 0.0002    | 0.0000**  | 0.0002    | 0.0000    |
|         | (0.9158)  | (2.0116)  | (1.4347)  | (1.6365)  |
| ListAge | 0.2952**  | 0.1732*   | 0.2295**  | 0.2747**  |
|         | (2.5799)  | (1.7670)  | (2.0628)  | (1.9949)  |
| FirmAge | 0.0123    | 0.0303    | 0.1141    | -0.0740   |
|         | (0.0748)  | (0.2429)  | (0.6973)  | (-0.4900) |
| Balance | 0.2878*   | -0.0488   | 0.1627    | 0.0941    |
|         | (1.8307)  | (-0.2717) | (1.1077)  | (0.4926)  |
| Dturn   | 0.0129    | 0.0357    | 0.0510    | -0.0317   |
|         | (0.2310)  | (0.5501)  | (1.1109)  | (-0.4479) |
| INST    | -0.3733   | -0.3186*  | -0.2993*  | -0.3613   |
|         | (-1.5731) | (-1.8098) | (-1.7479) | (-1.5620) |
| Mshare  | -0.3227   | 3.9878    | -0.2863   | -0.3279   |
|         | (-1.0046) | (0.3105)  | (-0.9834) | (-0.8599) |
| Mfee    | 0.5233    | 0.0169    | 0.8405    | 0.4658    |
|         | (0.6170)  | (0.0180)  | (1.1161)  | (0.4365)  |
| Occupy  | 0.9252    | -0.2212   | -0.4964   | 0.7550    |
|         | (0.5781)  | (-0.1783) | (-0.3065) | (0.5816)  |
| Big4    | -0.1141   | 0.1690*   | -0.0005   | 0.1019    |
|         | (-0.9006) | (1.9184)  | (-0.0039) | (1.0466)  |
| _cons   | 1.1941    | -1.3402   | -2.1741   | 0.9002    |
|         | (0.6973)  | (-0.6917) | (-1.1418) | (0.7110)  |
| Industry| Yes       | Yes       | Yes       | Yes       |
| Year    | Yes       | Yes       | Yes       | Yes       |
| N       | 3054      | 3026      | 2685      | 3395      |
| adj. $R^2$ | 0.125  | 0.113     | 0.114     | 0.092     |

*t* statistics based on robust standard error in parentheses $^{*} p < 0.1$, $^{**} p < 0.05$, $^{***} p < 0.01$

# 7 | CONCLUSIONS AND DISCUSSIONS

Using the panel data of Chinese listed companies, we construct a double difference model to analyze whether short selling will encourage enterprises to engage in green technology innovation activities, hypothesis 1 holds, and hypothesis 2 does not hold.

Our research results show that after the implementation of the margin trading and short selling policy, pilot companies will significantly increase their green technology innovation behavior. Although the existence of margin trading may lead to the management pursuing short-term effect and abandoning green technology innovation, the short selling threat effect brought by short selling mechanism is more powerful for promoting the quantity of green technology innovation, so we believe that the short selling threat and pressure brought by short selling to enterprises are the main reasons for enterprises to engage in green technology innovation. In addition, we believe that the margin trading and short selling mechanism is more similar to a supplementary mechanism spontaneously generated by the market.

The empirical results show that the implementation of margin trading and short selling will significantly promote the quantity of green technology innovation, but will not promote the quality of green technology innovation, we believe that the policy implementation can significantly improve the quantity of green innovation, but the effect on improving the quality of green innovation is not obvious. Possible reasons include: First, the short-term effect of shareholders and stakeholders, there are a large number of short-term investors in China's stock market, resulting in the restriction of stakeholders who lack long-term concept, this leads to the short-term innovation behavior that the management pays too much attention to the company's short-term economic interests rather than long-term social interests. Second, the margin trading brings the effect of management's participation in short-term speculation in order to maintain the stock price. Third, in order to meet the needs of stakeholders, the management may adopts the way of green camouflage to increase the number of technological innovation, so as to respond to stakeholder supervision at a relatively small cost. Fourth, environmental regulation and related policies did not pay attention to the assessment of green patent quality.

In addition, we found Several factors leading to the decline of the effect of policies on promoting the quantity of green technology innovation: The performance decline which increases the performance pressure, yield gap between financial assets and operating assets, the low risk of stock price collapse leads to low short trading volume and low short selling pressure, the increase of management shareholding and the decrease of institutional shareholding ratio lead to the decrease of external governance, the "crowding out effect" on innovation investment strategy caused by the intensification of product market competition, the decrease intensity of short selling transactions, the increase in the short-term effect of margin trading and the increase intensity of government formal environmental regulation.

Due to the dual effects of the margin trading policy, on the one hand, the incentive effect of the short selling mechanism can stimulate the green technology innovation of enterprises, on the other hand, the margin trading may cause the short-term behavior of enterprises, so it is necessary to consider and weigh the advantages and disadvantages in the implementation of the policy, try to formulate relevant supporting policies and guide the behavior of relevant investors to be rational and meet the needs of social development. If the combination of margin trading policy, relevant supporting policies (such as environmental policy, financial supervision policy and manager supervision mechanism) and investor behavior model can not inhibit the short-term behavior of enterprises, the actual effect of the policy may be greatly reduced, resulting in the increase of green technology innovation activities and quantity on the surface, and it can not produce high-level green innovation results and finally solve the environmental problems faced by mankind.

**APPENDIX A. VARIABLE DEFINITIONS**

This table contains the definitions of variables used in our analysis.

| | | |
|---|---|---|
| Dependent variable | $QGInnovation_{i,t}$ | Green patent quality of the company of company i in year t. |
| | $LN(NGInnovation_{i,t})$ | logarithm of number of green patents of the company of company i in year t. |
| Independent variable | $List_{i,t}$ | Whether the company is a policy pilot company. If yes, take 1; otherwise, take 0. |
| | $Policy2_{i,t}$ | Whether the company in the pilot province of carbon emission trading. If yes, take 1; otherwise, take 0. |
| Control variable | $RD_{i,t}$ | Ratio of R & D investment to total assets of company i in year t. |
| | $GDP_{i,t}$ | GDP of the province where the enterprise is located of company i in year t. |
| | $Lev_{i,t}$ | Total liabilities at the end of the year divided by total assets at the end of the year of company i in year t. |
| | $Growth_{i,t}$ | Current year's operating income / previous year's operating income - 1 of company i in year t. |
| | $Tanqi_{i,t}$ | Proportion of tangible assets, tangible assets / total assets of company i in year t. |
| | $Cfo_{i,t}$ | Enterprise cash flow, net cash flow from operating activities / total assets of company i in year t. |
| | $Board_{i,t}$ | The number of directors is taken as the natural logarithm of company i in year t. |
| | $Indep_{i,t}$ | Independent Directors divided by the number of directors of company i in year t. |
| | $Top1_{i,t}$ | Number of shares held by the largest shareholder / total shares of company i in year t. |
| | $Soe_{i,t}$ | The value of state-owned holding enterprise is 1, and that of others is 0 |
| | $ER_{i,t}$ | Proxy variable of environmental regulation, the ratio of provincial government environmental protection financial expenditure to GDP of company i in year t. |
| | $Roa_{i,t}$ | Net profit / average balance of total assets of company i in year t. |
| | $Roe_{i,t}$ | Net profit / average balance of shareholders' equity of company i in year t. |
| | $ATO_{i,t}$ | Operating income / average total assets of company i in year t. |
| | $Dual_{i,t}$ | If the chairman and the general manager are the same person of company i in year t, it is 1, otherwise it is 0 |
| | $BM_{i,t}$ | Book value / total market value of company i in year t. |
| | $TobinQ_{i,t}$ | (current stock market value + number of non tradable shares) × Net assets per share + book value of liabilities) / total assets of company i in year t. |
| | $Size_{i,t}$ | Total assets of the enterprise (100 million yuan) of company i in year t. |
| | $ListAge_{i,t}$ | Logarithm of (year of year - year of listing +1) of company i in year t. |
| | $FirmAge_{i,t}$ | Logarithm of (year of the current year - year of incorporation + 1) of company i in year t. |
| | $Balance_{i,t}$ | The sum of the shareholding ratio of the second to fifth largest shareholders divided by the shareholding ratio of the first largest shareholder of company i in year t. |
| | $Dturn_{i,t}$ | Average monthly stock turnover rate of the current year - average monthly stock turnover rate of last year of company i in year t. |
| | $INST_{i,t}$ | Total holdings of institutional investors divided by circulating share capital of company i in year t. |
| | $Mshare_{i,t}$ | Management shareholding data divided by total share capital of company i in year t. |
| | $Mfee_{i,t}$ | Administrative expenses divided by operating income of company i in year t. |
| | $Occupy_{i,t}$ | Capital occupied by major shareholders, other receivables divided by total assets of company i in year t. |
| | $Big4_{i,t}$ | The company is audited as 1 by four major companies (PWC, Deloitte, KPMG and Ernst & Young) of company i in year t., otherwise it is 0. |
| Other variables ( located in further study) | $EPS\_down_{i,t}$ | Whether the earnings per share of the current year decreased compared with last year of company i in year t, take 1 for the decline, otherwise take 0 |
| | $Short_{i,t}$ | the proportion of short selling balance to circulation market value of company i in year t. |
| | $Magrin_{i,t}$ | the proportion of margin trading balance to circulation market value of company i in year t. |
| | $Gap_{i,t}$ | Income gap between financial assets and operating assets of company i in year t. |
| | $DUVOL_{i,t}$ | The ratio of fluctuations in stock return of company i in year t |
| | $HHI_{i,t}$ | Herfindahl index, that is, the sum of squares of the proportion of the operating income of Companies to the operating income of all companies in the industry of company i in year t. |